**Title**

Transscleral Optical Phase Imaging of the Human Retina – TOPI.

**Authors**

T. Laforest[1,6]*†, M. Künzi[6]†, L. Kowalczuk[2,3], D. Carpentras[1], F. Behar-Cohen[2,4,5] and C. Moser[1]

**Affiliations**

[1] Laboratory of Applied Photonic Devices (LAPD), School of Engineering, École Polytechnique Fédérale de Lausanne (EPFL), Lausanne, Switzerland.

[2] Faculty of Biology and Medicine, University of Lausanne, Lausanne, Switzerland.

[3] Jules-Gonin Eye Hospital, Fondation Asile des aveugles, Lausanne, Switzerland.

[4] Centre de Recherche des Cordeliers, Sorbonne Université, Inserm, USPC, Université Paris Descartes, Université Paris Diderot, From physiopathology of ocular diseases to clinical developments, F-75006 Paris, France

[5] Ophtalmopole, Hôpital Cochin, Assistance Publique Hôpitaux de Paris, Université Paris Descartes, Paris, France.

[6] EarlySight SA, EPFL Innovation Park, Lausanne, Switzerland.

† Both authors contributed equally to this work.

* Corresponding author: timothe.laforest@epfl.ch

# Abstract

The *in vivo* observation of the human retina at the cellular level is crucial to detect lesions before irreversible visual loss occurs, to follow the time course of retinal diseases and to evaluate and monitor the early effects of treatments. Despite the phenomenal advances in optical coherence tomography (OCT) and adaptive optics systems, *in vivo* imaging of several retinal cells is still elusive.

Here we propose a radically different method compared to OCT, called transscleral optical phase imaging (TOPI), which allows to image retinal cells with high contrast, high resolution, and within an acquisition time suitable for clinical use. TOPI relies on high-angle oblique illumination of the retina, combined with adaptive optics, to enhance the phase contrast of transparent cells.

We first present *in-vivo* images of retinal cells, from the retinal pigment epithelium (RPE) to the nerve and vascular layers of the retina, in eleven healthy volunteers without pupil dilation. The morphology of the cells *in vivo* is then compared to that of images obtained with the same technique applied on *ex vivo* human RPE and pig retinas. Finally, we demonstrate the ability of high resolution phase microscopy to image pericytes and microglia around rat retinal capillaries.

Our results show the ability of TOPI to image and quantify retinal cells up to the RPE depth within a maximum time of 9 seconds over a field of view of 4.4 x 4.4°, opening new avenue in the *in vivo* exploration of the deepest layer of the retina in healthy and diseased eyes.

# Introduction

Retinal diseases are the major cause of blindness in industrialized countries. For example, an estimated total of 196 million people will be affected by age related macular degeneration in 2020 [1]. While tremendous effort is being made to develop novel therapeutic strategies to rescue retinal neurons and the retinal pigment epithelium, optimal means for evaluating the effects of such treatments are still missing.

Optical coherence tomography/microscopy (OCT/OCM) is a well-established depth-resolved imaging method used in ophthalmology. Together with scanning laser ophthalmoscope (SLO), they constitute the toolbox for routine eye fundus examination. However, minute changes in cell morphology which may be present during the early stages of disease progression cannot yet be quantified via commercial OCT or SLO systems. Therefore, the development of methods that can extract and quantify such small morphological changes in retinal cells at different depths is of considerable interest for new therapeutic approaches.

The retina is composed of many vascular and cellular layers, forming an intricate and complex tissue. Individual cell imaging in the retina is very challenging due to many constraints such as ocular aberration reducing lateral resolution, eye motion artifacts and, a lack of good contrast of transparent cells. Indeed, to ensure proper transmission of incident photons, the different layers laying anterior to the photoreceptors are well index-matched and transparent. On the other hand, the layers located posterior to the photoreceptors are difficult to visualize because most of the light is either absorbed or reflected at the photoreceptors interface, letting only few photons reaching the retinal pigment epithelium (RPE).

The use of adaptive optics (AO) to correct ocular aberration has enabled the imaging of the photoreceptor mosaic using both flood illumination [2] and scanning systems (AO-OCT, AO-SLO) [3] [4]. The contrast of the photoreceptor cells is obtained thanks to the high-reflection produced at the cell inner-outer segments interface [5] and posterior tip of the outer segment [6]. This relative high reflectivity is maximum for an illumination beam entering the eye pupil center and decreases sharply when entering close to the pupil edge making a larger illumination angle. The angular-dependent reflection of the retina has been extensively documented and is commonly known as optical Stiles-Crawford effect (SCE). SCE is due to the waveguiding properties of the photoreceptors that was first observed in the context of light perception by Stiles and Crawford in 1933 [7]. In recent works, optical methods to observe the neuroretina and the RPE cells were proposed using modified AO-SLO with offset aperture [8] [9] [10] or split detector setups [11] [12] [13]. However, due to the safety limitations in retinal illumination levels in humans, the signal-to-noise ratio of cell images was low [14]. In a recent study, Liu et al. [15] used an AO-OCT system to obtain high-resolution images of the neuroretina and the RPE. This optimized setup had enough resolution and sensitivity to detect signals from the organelles which were continuously moving within the cell soma. By averaging more than 100 images over a time span of approximately 10 minutes, *in vivo* images of ganglion somas and RPE cells were achieved. Despite the good quality of the images, this advanced OCT-AO technique is fundamentally limited by the biological time constant of organelle motility [16], which leads to an acquisition time that is still too slow for clinical use.

Here, we propose a method that circumvents the acquisition time limitation due to organelles motility and the contrast limitation resulting from trans-pupil illumination to provide cellular-resolution label-free high-contrast images of the retina layers. This method is based on transscleral flood illumination of the retina as illustrated in Fig. 1. Transscleral illumination has been used for decades for diagnostic in oncology and for retinal surgery but has never been used for cellular imaging [17]. The light transmitted through the sclera allows an oblique illumination of the posterior retina, which is then imaged using a trans-pupillary adaptive-optics full-field camera system. We call this imaging modality "TOPI" for Trans-scleral Optical Phase Imaging. As compared to scanning systems, this flood illumination technique is less sensitive to eye-motion artifacts since all the pixels are recorded simultaneously with a short acquisition time (few ms). Transscleral illumination has two interesting properties that compensate the low contrast of the retina observed with trans-pupil flood illumination. Firstly, due to the SCE, almost no high angle illumination light is coupled into the photoreceptors, allowing a large fraction of the illuminating light to reach the RPE layer. Secondly, no direct illumination light is collected by the imaging system because there is no overlap with the collection path. Only the light back-scattered by the different retina layers reaches the camera thus providing a darkfield imaging condition.

In this study, we first show that, by combining transscleral imaging and AO, our method allows visualization of RPE cells and other retinal structures at different areas of the fundus in eleven subjects. The potential of this technology is then validated *ex-vivo* using an experimental phase microscope with parameters similar to those of *in vivo* imaging. Finally, we present high resolution phase images of retinal capillaries taken under the phase microscope optimized for cell microscopy.

## Results

**Transscleral Optical Phase Imaging of the retina**

TOPI images were acquired using our experimental setup (Fig. S 1), in healthy volunteers, in dark room conditions without pupil dilation. The pupil diameter ranged from 4 to 7 mm. Images were taken at different depths, from the RPE to the inner retina.

***In vivo,* 4.4°×4.4 imaging of the retinal pigment epithelium within seconds**

The RPE layer from a healthy volunteer is exemplified in Fig. 2 (see Fig. S 2 for images in two additional volunteers). Due to the low signal over noise ratio (SNR) of the raw images coming from the retina, the *in vivo* TOPI image was processed using a recorded sequence of 100 images taken within a total acquisition time of 9 seconds. Due to the absorbing pigments in the cell soma, individual RPE cells are clearly visible on the full image with a dark center and bright edges. The low reflectivity of the cone mosaic caused by the high angle illumination is not shadowing the signal from the RPE cells underneath. Due to the large depth-of-field (i.e. ~20 μm), a lower spatial frequency bright-and-dark "cloudy" modulation is partially overlapping the RPE cellular mosaic. This unwanted signal is coming from the non-uniform pigmentation of the epithelium and from the out-of-focus light backscattered from the choroid located below. This modulation can be partially removed by post-processing the final image with a high-pass filter. The high-pass result is illustrated in Fig. 2B where the cell mosaic is easier to visualize.

Fig. 2A shows an example of the wide field-of-view image (4.4°×4.4°) obtained with the TOPI system in less than 10 seconds while providing cellular resolution. In comparison, an AO-OCT system of similar spatial resolution has a significantly smaller field of view of 1.5°x1.5° and takes several minutes to obtain a similar image [18].

**RPE cells size and density.**

RPE imaging was performed in eleven healthy volunteers (from 19 to 49 years-old). No major difference was noticed in the RPE images from the subjects with different skin and RPE pigmentations, demonstrating that TOPI could be used in patients of different phototypes [19]. The RPE cells quantification between 2° and 12° eccentricity gives an average area of $241 \pm 29$ μm$^2$ (Fig. 3A) for an average density of $4371 \pm 460$ cells/mm$^2$ (Fig. 3B).

Interestingly, the youngest participant presented the highest density of RPE cells, whereas the oldest participant presented the lowest density (Fig. 3C).

This quantitative analysis matched the values found in the literature at these eccentricities [15], [18]. One *in-vivo* autofluorescence AO-SLO study provided a RPE cell density of 5890 cells/mm$^2$ (168 µm$^2$ area) and 5630 cells/mm$^2$ (195 µm$^2$ area) at 5° and 6.25° respectively (2 subjects) [20]. Another AO-OCT study reported a mean density at 3° of 4975 ± 651 cells/mm$^2$ (6 subjects) [18]. From these three *in-vivo* measurements, no significant difference appears with respect to our data (*P>0.05* for the three comparisons). Scoles et al. reported, from histological study in the macula, an average density of 5662 cells/mm$^2$ and an average cell area of 160 µm$^2$ for age below 60 (6 samples) [21]. This density represents no significant difference with our data (*P=0.296*). Moreover, the row to row spacing of 13.8 µm extracted from Fourier analysis of the images (Fig. S 3) is the same as a previously reported value obtained with AO-OCT [18]. We also show on Fig. S 4B and C a stitched image of the RPE layer over 27° horizontally since our system allows imaging over a range of ±15° both vertically and horizontally (Fig. S 4A) . Finally, the SNR analysis presented in Fig. S 5 shows that an acquisition time as low as two seconds provides a suitable TOPI image for RPE quantification.

### *In vivo* imaging of the nerve fiber layer (NFL).

By axially shifting the focal plane of the imaging system, other layers of the neural retina can be observed, as shown in Fig. 4 on a 3°×3° cropped version of the full TOPI image for better visualization. The location of each crop is shown on the fundus image taken with an Optos SLO (Fig. 4A). Images of the vascular layer (Fig. 4B-C) and the nerve fiber layer (Fig. 4D-E-F) are shown. On the top row, retinal capillaries as small as few-microns can be seen and on the bottom row, the NFL at different eccentricities are imaged. Large axon bundles of 30-50 µm diameter are clearly visible and well-aligned in the direction of the optical nerve. The ability to observe very small capillaries and axons bundles comes from the dark-field effect of the transscleral illumination where the signal of the translucent structure is not buried under the illumination light.

**Comparison between transscleral and transpupillary illumination**

We then compared the commercial rtx1 AO retinal camera (Imagine Eyes), using flood transpupillary illumination, with our TOPI camera, using transscleral oblique illumination, in the same area of the subject S10's fundus as shown in the NFL images (Fig. 5A-D). The transpupillary illumination in rtx1 allows better visualization of the photoreceptors than TOPI images (Fig. 5B-E). This is due to the optical SCE at the inner-outer segment interface which provides a high reflectivity when light is incident perpendicular to the retina [22]. The strong cone signal produced by transpupillary illumination masks the adjacent layers, particularly the posterior RPE where only defocused photoreceptor light is visible (Fig. 5C). On the contrary, using transscleral illumination, cones produce a very weak signal, and thus the RPE cells appear well-contrasted (Fig. 5F). These results clearly confirm the very interesting properties of transscleral illumination for imaging the deepest layer of the retina.

***ex vivo* validation of the proposed phase imaging modality**

Our experimental phase microscope is illustrated in Fig. S 6. In a first series of experiments, it has been configured to obtain similar optical conditions to those of the *in-vivo* implementation in order to validate the potential of phase imaging with oblique illumination, without the artefacts induced by eye movements and aberrations. For this purpose, an objective with a numerical aperture (NA) of 0.25 was selected to match the maximum possible numerical aperture of a fully dilated human eye (0.24). Four light emitting diodes (LEDs) with a center wavelength of 650 nm and a 50 nm bandwidth were used to illuminate the retina. The *ex-vivo* image reconstruction is described in the Materials and Methods section. To mimic the scattering effect of the sclera, a white paper sheet was placed between the LEDs and the samples (Fig. S 6D).

**Optical characterization of *ex-vivo* images**

We first measured the modulation transfer function (MTF) resulting from the incoherent illumination (Fig. S 7A). Then, we performed imaging of the frontal frozen sections of a pig neuroretina to compare our phase imaging method with digital holographic microscopy (DHM). For phase imaging, a scattering surface (standard printer white paper) was attached to the bottom side of the glass slide in order to simulate the back-scattered light (Fig. S 6B). DHMs produce true metrological phase images, allowing for a quantitative comparison with the proposed method Fig. S 7B). The images from the DHM were taken with a 0.4-NA microscope objective and with coherent light illumination. Because our method uses incoherent light, the cut-off spatial frequency was equal to $2NA/\lambda = 0.5/\lambda \ [m^{-1}]$, with $\lambda$ being the central illumination wavelength. Thus, our phase imaging method produces images with a resolution similar to the 0.4-NA DHM images taken with coherent illumination and with a sharp cut-off spatial frequency of $NA/\lambda = 0.4/\lambda \ [m^{-1}]$.

Finally, we compared our phase imaging modality with reflectance confocal microscopy using a human retina-choroid complex. The back illumination was provided by the scattering of the RPE-choroid layers (Fig. S 6C), and the region of interest was selected using blood vessels as landmarks. Fig. S 7C shows the intensity images taken with a confocal microscope and 0.8-NA (left) and 0.3-NA (center) objectives. Our phase image (right), which was recorded at the same depth, exhibits better SNR and resolution than the 0.3-NA confocal microscope image. In addition, compared with the 0.8-NA image, the phase image makes other features visible, such as blood cells into the vessels and other retinal cellular structures outside of the vessels.

### Validation of phase imaging of retinal layers

A second healthy human retina-choroid complex was observed *ex vivo* to validate, under steady conditions, the morphology of the RPE cell imaged in healthy volunteers. The microscope was set with one LED to mimic the *in vivo* set up and with a 0.17 NA objective to mimic a pupil without dilation. As in *in vivo* phase images, RPE cells appeared with a dark center and bright edges in the *ex vivo* dark-field images of the human RPE (Fig. S 8). This feature has also been observed in dark-field AO-SLO [23].

The phase microscope set with four LEDs and the 0.25 NA objective lens was then used to demonstrate the depth-resolution of our phase modality. Depth scanning was achieved by moving the sample axially, following the z-axis. Thanks to the objective's 10 µm depth of field, all layers of the 200-µm-thick neuroretina on a pig retina-choroid complex were visualized in the *area centralis* [24] (Fig. 6A, C, E, G, I, K). The raw phase images were processed in order to digitally tag cell nuclei in each layer (Fig. 6B, D, F, H, J, L), and thus to quantify cell densities in the ganglion cell layer (2'260 cells/mm$^2$), the inner nuclear layer (9'744 cells/mm$^2$), the outer nuclear layer (23'790 cells/mm$^2$), and the photoreceptor layer (20'930 cells/mm$^2$) (Fig. 6M, blue curves). These results were compared with the data available in the literature, and we found that the measured ganglion cell and cone densities are consistent with the reported ganglion cells (1500 to 4000 cells/mm$^2$, [25]) and cones (19000 [24] to 22600 [26] cells/mm$^2$) densities.

**High resolution phase microscopy of retinal capillaries**

Finally, our experimental phase microscope offered the opportunity to perform high resolution phase imaging in rat retinal flatmounts. For this purpose, we observed retinal capillaries and their surrounding cells, using microscope objectives with higher NA, in a correlative study between phase microscopy and fluorescence confocal microscopy.

The retina from a healthy rat was first immuno-stained for NG2 to specifically detect pericytes in fluorescence. Using vessels as landmark, the same region was observed with confocal microscopy and with our experimental phase microscope set with a 0.45-NA objective (Fig. 7A-C). The comparisons of these images demonstrated that the proposed phase imaging method allows for the detection of pericytes, appearing as bright cells around retinal capillaries (Fig. 7A) exactly at the same location of NG2+ cells (Fig. 7B). The counter-staining with DAPI (Fig. 7C) demonstrated that phase microscopy also allows for the detection of the nuclei cells in the observed layer.

Then, we observed the retina from an 11-month-old rat presenting retinal dystrophy associated with telangiectasias [27], using a 0.65NA objective in the phase microscope (Fig. 7D) after the immuno-staining of Iba1, the marker of microglial cells (Fig. 7E, green), and of collagen IV to highlight vessels (Fig. 7E, red). The image correlation analysis between phase and confocal microscopy unequivocally demonstrated that high resolution phase imaging allows for the detection of these small cells as well.

**Discussion**

Currently, the standard of care for retinal imaging is based on OCT and SLO. The first commercially available device integrating AO was based on flood illumination (rtx1™, Imagine Eyes). We describe herein a novel microscopy concept and instrument for fast *in vivo* imaging of the retinal pigment epithelium with high contrast and cellular resolution.

To our knowledge, no other imaging system allows for the visualization of RPE cells within a few seconds (Table 1). We see no technological barriers to create an easy-to-use clinical instrument because transscleral illumination requires the same accuracy in terms of beam alignment than transpupillary systems. On the contrary, for cellular-resolution imaging, AO SLO and AO OCT would be more difficult to implement in clinical setting for the following technical reasons. As presented in Table 1, their field of view of 2°×2° is close to five times smaller than the current field of view of the TOPI system. Since a large area of observation is more clinically relevant, five times more images would be required. Moreover, the use of a non-scanning system makes the acquisition process more stable under the effects of eye motion, especially because AO-SLO and AO-OCT require measurement time of several minutes (Table 1). For the latter, image distortion becomes a problem, which is less the case for a flood-illumination system, in which the full field of view is recorded simultaneously.

The speed of an eye examination is crucial for patient's comfort but more importantly when fixation capacity is reduced by retinal disease. The typical scan time for commercial clinical devices is a few seconds (2–3 s for Canon OCT-HS100, 2 s for Imagine Eyes rtx1, 3 s for Optovue AngioVue, 19–38 ms for B-Scan for Heidelberg Spectralis OCT, and 2 s for Zeiss Cirrus HD-OCT). With less than ten seconds required for TOPI acquisition, the obtained contrast was sufficient to perform quantitative analyses on RPE cells. The full field flood imaging technique combined with transscleral illumination establishes an imaging modality that is several orders of magnitude faster than the AO-OCT technique when applied to weakly reflecting cells [15], [16]. This difference is essentially due to the fact that TOPI uses the scattered light from the back of the eye, which is much more reflective than the transparent cells.

The oblique flood illumination enables fast full-field image acquisition that does not depend on biological processes, such as organelle motion [18], [15]. The light radiant exposure at the imaged area of the retina that is comparable with existing technologies. Fig. S 9 shows the estimated retina radiant exposure level of competing AO imaging methods. For TOPI imaging, a single raw image requires an estimated radiant energy of 79 µJ/cm$^2$ at the retina, while the 100-images sequence needed to obtain the final image yields 7.9 mJ/ cm$^2$. These values are comparable to those of the commercial flood illumination device rtx1$^{TM}$ [2] and are lower than those used in a scanning system based on SLO or OCT. The projected spot on the sclera has a power density that satisfies the safety standard for skin and ocular tissues (see the Materials and Methods section). We note that, in all previous studies related to high-resolution imaging of the retina, averaging of raw data is also necessary to obtain high SNR.

In the present study, *in vivo* examinations of eleven healthy volunteers with our TOPI set up demonstrated the feasibility of fast imaging of RPE cells but also the possibility of quantifying them accurately. Importantly, the quality of RPE images was not impaired by individual skin phototypes and subsequent RPE pigmentation. TOPI examination also allowed accurate visualization of nerve and vascular fiber layers. Compared to images of retinal capillaries obtained with OCT angiography, the images of the smallest vasculature obtained with our method are independent of flow velocity and could thus provide complementary information.

Using an experimental microscope, the application of our phase imaging method to the observation of post-mortem human RPE validated the morphology of the RPE cells observed in volunteers. We have also shown that this imaging capability can be used to visualize all the layers of a pig neuroretina by combining multiple oblique illumination points and with appropriate modeling of the illumination function. Finally, the optimization of this setup for high resolution microscopy has shown that this new modality allows the visualization of the retinal capillaries and their surrounding cells, pericytes and microglia, without immuno-staining.

The ability to image RPE cells within seconds in patients is of major interest to better understand the physiopathogenesis of retinal diseases and to follow the course of diseases that affect the RPE such as age-related macular degeneration (AMD), retinitis pigmentosa [28] [29], or diabetic retinopathy [30]. Decrease in cell density and change in cell morphology could potentially be detectable at the early stage of AMD, before any complication has occurred [31] [32]. TOPI could also be used to detect effects of drugs and treatments at a cellular level, offering the opportunity to intervene before irreversible functional loss has occurred. Morphologic quantitative new endpoints could thus emerge. In addition, recent medical studies have shown that Parkinson's and Alzheimer's diseases have an effect on the neuronal layers of the retina, [33] [34], which may increase the impact of the clinical use of this technology.

In summary, TOPI holds promise for clinical use because it is, to the best of our knowledge, the only technique fast enough to image the RPE with high contrast and cellular resolution. Further clinical studies will clarify the limitations of this new imaging modality particularly in diseased states that distort the retinal layers organization. TOPI could be combined with OCT methods, enabling not only to accelerate our understanding of retinal diseases and to define new morphological endpoints for therapeutic evaluation, but also the understanding of other neuro degenerative diseases.

## Materials and Methods

**Study on human subjects.**

The single-center study aimed at validating the performances of our phase imaging device adheres to the tenets of the Declaration of Helsinki. The protocol involving healthy human volunteer subjects was approved by the local Ethics Committee of the Swiss Department of Health on research involving human subjects (CER-VD N°2017-00976). Informed consent was obtained from the subjects. The left eyes of eleven healthy volunteers were imaged (four women, seven men) with an average age of 28 years (+/− 9) and having different skin pigmentations ranging from 1 to 6 on the Fitzpatrick scale were imaged [19].

**Transscleral illumination principle**

Transscleral illumination of the fundus produces an oblique beam that reaches the posterior layers of the eye. The sclera transmits about 30% of the incident light at the imaging wavelength of 810 nm [17], and the absorption and reflectance of the RPE-choroid complex vary with its pigmentation [35] [36]. While crossing the sclera, the illumination light is scattered and diffused, producing a uniform highly-divergent beam that reaches obliquely the posterior retina and illuminates it evenly. The back-scattered light coming from all the retina layers is then captured through the pupil by the optical system. The achieved illumination angle is thus larger than what is obtainable via illumination through the pupil. We showed in a previous study that it creates a non-uniform excitation of the retinal spatial frequencies and enhances the contrast of phase objects [37] [38]. Additionally, the use of partially coherent light yields a system bandwidth that is twice the bandwidth obtainable using coherent light.

*In vivo* **experimental design**

For the in vivo demonstration, the oblique illumination of the retina was produced using two Light Emitting Diodes (LED) with 810 nm nominal wavelength, 25 nm FWHM. The LED light beams are focused on the sclera at the pars-plana region to the temporal and nasal side of the iris (Fig. S 1). The light beams can be operated at the same time or alternatively to capture images with one or two illumination points. The obtained angle of illumination for each point was therefore approximately 35°. The subject's head and eye were aligned with the system using an adjustable chin and forehead rest.

The optical setup shown in Fig. S 1 consists in the following elements. The illumination arm was composed of a yellow fixation target, of a 756-nm superluminescent diode (SLD, Exalos), which provided a point source at the retinal plane for the aberration correction measurements, and of an 1050 nm low power infrared LED for illuminating the pupil uniformly. The imaging arm integrated a pupil camera, an AO feedback loop [39] (a Imagine Eyes Mirao 52e or Alpao DM97-15 deformable mirror and Imagine Eyes HASO4 wavefront sensor), and an sCMOS camera for imaging the retina (ORCA Flash 4.0, Hamamatsu). The retinal sCMOS camera collected the light coming from the transscleral illumination to image the eye fundus through the pupil with a lateral sampling of 0.7 µm/pixel.

Once the position of the eye was aligned on the setup thanks to a chin and forehead rest, the retinal camera, which was mounted on a translation stage, was adjusted to focus on the targeted retinal layer. The real-time AO feedback loop was used for correcting the defocus accommodation of the eye (i.e., axial shifts of the plane), thus always locking a given imaging depth in the retina to a corresponding camera position. The depth sectioning is limited in this case by the depth of field, which is given by the pupil size and the aberrations of the eye. For instance, for an aberration-free-6-mm pupil, the depth of field is 27 µm, yielding a value of 0.17 NA (the equivalent focal distance of the eye lens in air is 17 mm), for a lateral resolution of 2.9 µm. A sequence of one hundred images at a given depth in the retina was acquired by successively turning the LED spots ON and OFF with an exposure time of 8 ms and a repetition rate of 11 Hz per illumination point. The image focus can be made at any depth in the retina by introducing a quadratic defocus term on the deformable mirror. Each layer took 9 seconds to image. A hundred raw images for each illumination point were captured, for a total of 200 raw images for a single depth. For the RPE, an analysis based on its amplitude spectrum showed the possibility of obtaining a good SNR with much fewer raw images (2 seconds would be sufficient, as illustrated in Fig. S 5). The raw images were then post-processed, as explained in the "in-vivo image processing" section, to produce a single TOPI image.

**Light safety analysis**

The illumination beams are projected on the participant's sclera when the eye is correctly aligned with the instrument. However, a complete safety analysis has to take into account any accidental projection of the beam on the skin, iris, cornea and lens to ensure no potential hazard of the system. To this end, the scleral, corneal, lens and retinal exposure were measured or calculated and compared with their respective exposure limits. We used the limits as expressed in the international standards for ophthalmic instrument optical safety ISO 15004-2:2007 [40] for the ocular tissue and the skin limits given in the 2013 ICNIRP Guidelines on limits of exposure to incoherent visible and infrared radiation [41].

**Eye exposure conditions:**

The beam minimum diameter is 3.5 mm at its focus point on the eye and the full converging angle is 0.37 radian. The pulse length is 8 ms and the repetition rate 11 Hz (duty cycle of 0.088), with a measured pulse power of 250 mW at a center wavelength of 810 nm with FWHM of 25 nm. The consecutive acquisition time is no longer than 10 seconds. Based on the measured power and pulse specification, the eye exposure is straightforward calculated:

Maximum pulsed irradiance at the sclera surface (8ms): $E_P = 2.598$ W/cm$^2$.

Scleral radiant exposure for single pulse (1 pulse, 8ms): $H_1 = 21$ mJ/cm$^2$

Time-average irradiance (11 Hz): $E_{AV} = 229$ mW/cm$^2$.

Scleral radiant exposure for maximum exposure time (110 pulses, 10s): $H_{110} = 2.287$ J/cm$^2$

**Retinal exposure:**

The light exposure reaching the underlying choroid and retina after crossing the sclera is less straightforward to compute. Only part of the light will transmit, and the beam will be scattered and spread by the tissue. The more accurate measurement on scleral transmission and absorption were done by Vogel et al. in 1991 [17]. They measured a ~33% transmission (and ~5% absorption) for a wavelength of 810 nm. As a conservative estimate, we will assume no diffusion nor beam spreading, thus giving a retinal exposure of 33% of the scleral exposure to compare with the limits.

The retinal irradiance in case of accidental trans-pupil illumination is calculated thanks to the beam converging angle. The retinal image diameter $d_r$ and surface $A$ is:

$d_r = f \cdot \varphi$ = (17 mm)(0.37 radian) = 6.29 mm → $A = 0.31$ cm$^2$

The irradiance is then directly found: $E$ = 250 mW/0.31 cm$^2$ = 0.805 W/cm$^2$.

The irradiance of the posterior retina in case of accidental trans-pupil exposure (0.805 W/cm$^2$) is smaller than the irradiance of the retina underlying the sclera in case of intended transscleral exposure (33% * 2.598 = 0.857 W/cm$^2$). The transscleral exposure and not the accidental trans-pupil exposure is therefore the worst-case exposure scenario for the retina to compare with the limits.

**Applicable limits for Group 1 pulsed instrument [40]:**

Weighted retinal visible and infrared radiation radiant exposure limit:

For 1 pulse (8ms): $H_{VIR-R}$ = 0.1605 J/cm$^2$

For 110 pulses (10s): $H_{VIR-R}$ = 10.4184 J/cm$^2$

Unweighted corneal and lenticular infrared radiation radiant exposure limit:

For 1 pulse (8ms): $H_{IR-CL}$ = 0.5383 J/cm$^2$

For 110 pulses (10s): $H_{IR-CL}$ = 3.2009 J/cm$^2$

Unweighted anterior segment visible and infrared radiation radiant exposure limit (for convergent beam only):

For 1 pulse (8ms): $H_{VIR-AS}$ = 7.4767 J/cm$^2$

For 110 pulses (10s): $H_{VIR-AS}$ = 44.4570 J/cm$^2$

**Skin limits [41]:**

For 1 pulse (8ms): $H_{SKIN}$ = 0.5981 J/cm$^2$

For 110 pulses (10s): $H_{SKIN}$ = 3.5566 J/cm$^2$

The radiant exposure conditions of the eye for 1 pulse $H_1$ and 110 pulses $H_{110}$ calculated above can be directly compared with the different limits. They are below all limits for Group 1 instrument (i.e. "ophthalmic instrument for which no potential light hazard exists" [40]) as well as below the ICNIRP skin limits [41]. The retina limits for our NIR exposure are not even attained if the sclera were considered with 100% transmission. The safety factor for retinal exposure is therefore at least a factor 20 (considering 33% transmission and thermal hazard weighting function R($\lambda$)=0.6 for $\lambda$=810 nm).

The condition being the closest to the limit is the cornea and lenticular exposure in case of misalignment of the device, or a subject starring directly at the illumination beam. This event is very unlikely to happen but a safety factor of 1.5 prevent any potential adverse effect in that case.

No explicit limits are provided for the sclera in the ISO 15004-2:2007 or ICNIRP guidelines. However, limits are provided for the anterior segment (cornea and lens) and for the skin. The sclera tissue itself can be considered in between skin and cornea. As it has a similar absorption coefficient for NIR as the cornea (but more scattering) [17], the sclera is expected to have a similar exposure limit. It is therefore reasonable to assume that sclera is safe if both the anterior segment limits and the limits for the more pigmented tissues such as skin are not attained, which it is the case for our instrument. As an additional safety evidence, far larger radiant exposure was sent on the sclera by Geffen et al. without noticing any adverse effect [42].

**In vivo image processing**

Because of sudden eye micro-saccades during the 8ms acquisition time, some raw images had motion-blur artifacts. Therefore, only the 80% sharpest images were automatically or manually selected for averaging at each illumination point. The averaging of raw images was carried out to increase the SNR. Before averaging, the images were aligned in order to correct for the eye motions using the ImageJ plugin TurboReg, which considers the translations and the rotations of the images [43]. Once the images were aligned, they were then averaged, high-pass filtered (ImageJ bandpass filter with boundaries 1px-30px) and normalized with histogram stretching. After averaging, a single high-SNR dark-field image was obtained for each illumination point for a specific layer. The high-pass filter was used to remove a large part of the out-of-focus light.

**Cell segmentation and quantification**

Cell segmentation was performed on several retinal layers to obtain quantitative morphology and density values. For the RPE layer, an automatic segmentation was performed via local minimum detection. Then, the Voronoi filter function was applied to create equidistant cell contours between the two nearest maxima, leading to RPE segmentation. Voronoi mapping is commonly used to map the retinal soma mosaic [15]. The coefficient of variation (CoV) [44], has been computed on subject S10 based on automatic RPE count and area measurements repeated five times under changed equipment and materials (two different deformable mirrors and two different head mounts were used) over five month. The results, plotted on Fig. S 10, give a CoV of 6.13% for RPE count and 6.44% for RPE area. This standard measurement characterizes the reproducibility of the acquisition. The resulting values are considered good for a prototype device since higher values of CoV can be measured on commercial devices [44].

**Statistical analysis**

Results were expressed as mean ± standard deviation. Student's t-test was used to compare our data with that of the literature, when possible. A $P$ value of <0.05 was considered statistically significant.

**Retinal *ex vivo* samples**

**Flat-mounted retina-choroid complexes**

The two human-donor eyes were obtained from the eye bank of the Jules-Gonin Eye Hospital, in conformity with the Swiss Federal law on transplantation. It was collected for research purpose at 10 h post mortem, fixed in a 4% paraformaldehyde solution (PFA) for 24 h, and then stored in 1% PFA at +4°C until it was analyzed. Pig eyes were obtained from a local slaughterhouse, a few hours after the animals' death. These eyes were fixed in 4% PFA for 24 h, and then stored in a phosphate-buffered saline (PBS) h at +4°C for one day before mounting.

After washing in PBS, the anterior segment of the eyes and the vitreous were removed. The retina-choroid complexes were detached from the remaining posterior segments, and flat-mounted in a solution made of PBS-glycerol (1:1). The human retina-choroid complexes were observed using a confocal microscope (Zeiss LSM710). The human and pig samples were also observed using our experimental phase microscope (as illustrated in Fig. S 6D).

### Cryosectioning

One pig eye was kept for cryosectioning. After fixation and microdissection, the posterior segment of the eye was snap- frozen prior to frozen sectioning on a microtome-cryostat. Frontal 10 um-thick sections were prepared and mounted on a glass coverslip. These sections were imaged using a digital holographic microscope (DHM) [45] and then with our phase microscope.

### Immunohistochemistry on neuroretinas

Two pigmented rats from the animal facility of the Jules-Gonin Eye Hospital were used. Investigations were performed in accordance of the ARVO statement for the Use of Animals in Ophthalmic Vision Research. Our study was approved by the cantonal veterinary office (Authorization VD2928). Retinal morphology was examined *in vivo*, under ketamine / xylazine (80mg/kg and 8 mg/kg) anesthesia and after dilation, using a spectral domain OCT system adapted for rat eyes (Bioptigen). Animals were sacrificed via pentobarbital injection. After enucleation and fixation of the whole eyes during 2 h in 4% PFA, the anterior segments were discarded and the neuroretinas were carefully separated from the remaining posterior segment. Post-fixation was carried out for 10 min at −20°C in acetone.

For pericytes immunostaining, after rehydration in PBS supplemented with 0.5% Triton X100 and 10% fetal bovine serum (FBS) overnight at +4°C, the retinas were incubated for two days at +4°C with polyclonal rabbit anti-NG2 antibodies (1:200; AB5320, Merk Millipore). For microglia immunostaining, after rehydration in PBS supplemented with 0.5% Triton X100 and 10% FBS for 30 minutes at room temperature, the retinas were incubated overnight at +4 °C with a rabbit anti-iba1 (1:200; Wako) and a goat anti-collagen IV (1:100, Serotec) antibodies diluted in the same buffer. After washing, the neuroretinas were incubated with the appropriate antibody (Alexa-488-conjugated goat anti rabbit IgG, 1:250; donkey anti-goat A633, 1:400; Invitrogen), diluted in PBS supplemented with 0.1% Triton X100 and 10% FBS, for 2 h at room temperature. After washing, the tissues were stained for 10 min with 4', 6-Diamidino-2-Phenyl-Indole (DAPI, 1:10 000), washed again, and flat-mounted in PBS-glycerol (1:1). The neuroretinas were examined with a confocal microscope (Zeiss LSM 710), and then with our phase microscope.

## *Ex vivo* experimental design

Our experimental phase microscope is illustrated in Fig. S 6. A total of four different LEDs were placed equally spaced around a microscope objective to produce four different illumination points. The sample was placed below the microscope objective (MO) and imaged using a CMOS camera using the MO and an achromatic doublet as a tube lens. A scattering layer of translucent commercial tape was used between each LED and the sample to increase the extent of illumination light and to mimic the scattering caused by the sclera. The sample was moved vertically to change the imaging layer depth. Depending on the NA of the MO, the exposure time varied between 1 ms and 20 ms. For each illumination point, a total of ten raw images were averaged to increase the SNR. A total of four averaged dark-field images (one for each points) were obtained for each depth. Then, the processing steps described below were carried out to produce either differential phase contrast (DPC) images or a single phase-contrast image.

### *Ex vivo* image processing and phase reconstruction

An image of the sample is recorded for each illumination point and a final quantitative phase image is reconstructed following the method described in [38]. Phase images can be reconstructed using the following principle: at least two pictures with intensities I(α) and I(α+180), captured with two different symmetric illumination polar angles (α and α+180°), are required. A DPC image can be then obtained according to:

$$I_{DPC} = \frac{I(\alpha) - I(\alpha + 180°)}{I(\alpha) + I(\alpha + 180°)} \ .$$

Using several illumination points, multiple DPC images are produced. By combining these different DPC images of the same object, it is possible to obtain a quantitative phase image of the object thanks to the reconstruction process described in [46]. Finally, the reconstruction process of the phase is performed [47] [46] [38].

The reconstruction process is a Tikhonov regularization process with a known transfer function. The absorption transfer function is supposed to be identical for the different illumination points, meaning that it cancels out after the subtraction. The phase transfer function was calculated using the linearized imaging model presented by Tian and Waller [46].

### Code availability

The code for performing the reconstruction process may be requested from the authors.

# References

[1]  W. L. Wong, X. Su, X. Li, C. M. G. Cheung, R. Klein, C.-Y. Cheng and T. Y. Wong, "Global prevalence of age-related macular degeneration and disease burden projection for 2020 and 2040: a systematic review and meta-analysis," *The Lancet Global Health,* vol. 2, no. 2, pp. e106-e116, 2014.

[2]  C. Viard, K. Nakashima, B. Lamory, M. Pâques, X. Levecq and N. Château, "Imaging microscopic structures in pathological retinas using a flood illumination adaptive optics retinal camera," *Ophthalmic Technologies XXI,* p. 788501, 17 February 2011.

[3]  E. J. Fernández, I. Iglesias and P. Artal, "Closed-loop adaptive optics in the human eye," *Opt. Lett.,* vol. 26, no. 10, pp. 746-748, 2001.

[4]  P. Artal and R. Navarro, "High-resolution imaging of the living human fovea: measurement of the intercenter cone distance by speckle interferometry," *Opt. Lett.,* vol. 14, no. 20, pp. 1098-1100, 1989.

[5]  W. D. H. H. Pallikaris A., "The Reflectance of Single Cones in the Living Human Eye," *Invest Ophthalmol Vis Sci,* vol. 44, no. 10, pp. 4580-92, 2003.

[6]  W. Gao, B. Cense, Y. Zhang, R. S. Jonnal and D. T. Miller, "Measuring retinal contributions to the optical Stiles-Crawford effect with optical coherence tomography," *Optics Express,* vol. 16, no. 9, pp. 6486-6501, 2008.

[7]  W. S. Stiles and B. H. Crawford, "The Luminous Efficiency of Rays entering the Eye Pupil at Different Points," *Proceedings of the Royal Society of London,* vol. 112, no. 778, pp. 428-450, 1933.

[8]  A. Guevara-Torres, D. R. Williams and J. Schallek, "Imaging translucent cell bodies in the living mouse retina without contrast agents," *Biomedical Optics Express,* vol. 6, no. 6, pp. 2106-2119, 2015.

[9]  T. Y. Chui, T. J. Gast and S. A. Burns, "Imaging of vascular wall fine structure in human retina using adaptive optics scanning laser ophthalmoscopy," *Investigative Ophthalmology & Visual Science,* vol. 54, no. 10, pp. 7115-7124, 2013.

[10] T. Y. P. Chui, D. A. Van Nasdale and S. A. Burns, "The use of forward scatter to improve retinal vascular imaging with an adaptive optics scanning laser ophthalmoscope," *Biomedical Optics Express,* vol. 3, no. 10, pp. 2537-2549, 2012.

[11] W. B. Amos, S. Reichelt, D. M. Cattermole and J. Laufer, "Re-evaluation of differential phase contrast (DPC) in a scanning laser microscope using a split detector as an alternative to differential interference contrast (DIC) optics," *Journal of Microscopy,* vol. 210, no. 2, pp. 166-175, 2003.

[12] D. Cunefare, R. F. Cooper, B. P. Higgins, A. Dubra, J. Carroll and S. Farsiu, "Automated Detection of Cone Photoreceptors in Split Detector Adaptive Optics Scanning Light Ophthalmoscope Images," *Investigative Ophthalmology & Visual Science,* vol. 57, no. 12, pp. 61-61, 2016.

[13] Y. N. Sulai, D. Scoles, Z. Harvey and A. Dubra, "Visualization of retinal vascular structure and perfusion with a nonconfocal adaptive optics scanning light ophthalmoscope.," *Journal of The Optical Society of America A-optics Image Science and Vision,* vol. 31, no. 3, pp. 569-579, 2014.

[14] E. A. Rossi, C. E. Granger, R. Sharma, Q. Yang, K. Saito, C. Schwarz, S. Walters, K. Nozato, J. Zhang, T. Kawakami, W. Fisher, L. R. Latchney, J. J. Hunter, M. M. Chung and D. R. Wiliams, "Imaging individual neurons in the retinal ganglion cell layer of the living eye," *PNAS,* vol. 114, no. 3, pp. 586-591, 2017.

[15] Z. Liu, K. Kurokawa, F. Zhang, J. J. Lee and D. T. Miller, "Imaging and quantifying ganglion cells and other transparent neurons in the living human retina," *Proceedings of the National Academy of Sciences of the United States of America,* vol. 114, no. 48, pp. 12803-12808, 2017.

[16] Z. Liu, K. Kurokawa, F. Zhang and D. T. Miller, "Characterizing motility dynamics in human RPE cells," *Proc. of SPIE,* pp. 1004515-1-7, 8 February 2017.

[17] A. . Vogel, C. . Dlugos, R. . Nuffer and R. . Birngruber, "Optical properties of human sclera, and their consequences for transscleral laser applications," *Lasers in Surgery and Medicine,* vol. 11, no. 4, pp. 331-340, 1991.

[18] Z. . Liu, O. P. Kocaoglu and D. T. Miller, "3D Imaging of Retinal Pigment Epithelial Cells in the Living Human Retina.," *Investigative Ophthalmology & Visual Science,* vol. 57, no. 9, p. , 2016.

[19] T. B. Fitzpatrick, "The Validity and Practicality of Sun-Reactive Skin Types I Through VI," *Arch Dermatol.,* vol. 6, no. 126, pp. 869-871, 1988.

[20] J. I. W. Morgan, A. Dubra, R. Wolfe, W. H. Merigan and D. R. Williams, "In Vivo Autofluorescence Imaging of the Human and Macaque Retinal Pigment Epithelial Cell Mosaic," *Invest. Ophthalmol. Vis. Sci.,* vol. 50, no. 3, pp. 1350-1359, 2009.

[21] S. K. Bhatia, A. Rashid, M. A. Chrenek, Q. Zhang, B. B. Bruce, M. Klein, J. H. B. Boatright, Y. Jiang, H. E. Grossniklaus and J. M. Nickerson, "Analysis of RPE morphometry in human eyes," *Molecular Vision,* vol. 22, pp. 898-916, 2016.

[22] M. . Zacharria, B. . Lamory and N. . Chateau, "Biomedical imaging: New view of the eye," *Nature Photonics,* vol. 5, no. 1, pp. 24-26, 2011.

[23] D. Scoles, Y. N. Sulai and A. Dubra, "In vivo dark-field imaging of the retinal pigment epithelium cell mosaic.," *Biomedical Optics Express,* vol. 4, no. 9, pp. 1710-1723, 2013.

[24] Chandler, Smith, Samuelson and MacKay, "Photoreceptor density of the domestic pig retina," *Veterinary Ophthalmology,* vol. 2, no. 3, pp. 179-184, 1999.

[25] M. Garcá, J. Ruiz-Ederra, H. Hernández-Barbáchano and E. Vecino, "Topography of pig retinal ganglion cells," *The Journal of Comparative Neurology,* vol. 486, no. 4, pp. 361-372, 2005.

[26] C. G. Gerke, Y. Hao and F. Wong, "Topography of rods and cones in the retina of the domestic pig," *Hong Kong Medical Journal,* vol. 1, pp. 302-308, 1995.

[27] M. Zhao, C. Andrieu-Soler, L. Kowalczuk, M. Paz Cortés, M. Berdugo, M. Dernigoghossian, F. Halili, J. Jeanny, B. Goldenberg, M. Savoldelli, M. El Sanharawi, M. Naud, W. van Ijcken, R. Pescini-Gobert, D. Martinet, A. Maass, J. Wijnholds, P. Crisanti, C. Rivolta and F. Behar-Cohen, "A new CRB1 rat mutation links Müller glial cells to retinal telangiectasia.," *J Neurosci.,* vol. 35, no. 15, pp. 6093-6106, 2015.

[28] I. Bhutto and G. Lutty, "Understanding age-related macular degeneration (AMD): relationships between the photoreceptor/retinal pigment epithelium/Bruch's membrane/choriocapillaris complex.," *Mol Aspects Med.,* vol. 33, no. 4, pp. 295-317, 2018.

[29] M. Humayun, M. Prince, E. J. de Juan, Y. Barron, M. Moskowitz, I. Klock and A. Milam, "Morphometric analysis of the extramacular retina from postmortem eyes with retinitis pigmentosa.," *Invest Ophthalmol Vis Sci.,* vol. 40, no. 1, pp. 143-148, 1999.

[30] P. Rothschild, S. Salah, M. Berdugo, E. Gélizé, K. Delaunay, M. Naud, C. Klein, A. Moulin, M. Savoldelli, C. Bergin, J. Jeanny, L. Jonet, Y. Arsenijevic, F. Behar-Cohen and P. Crisanti, "ROCK-1 mediates diabetes-induced retinal pigment epithelial and endothelial cell blebbing: Contribution to diabetic retinopathy.," *Sci Rep.,* vol. 7, no. 1, p. 8834, 2017.

[31] F. C. Ferrington D.A., "Perspective on AMD Pathobiology: A Bioenergetic Crisis in the RPE.," *Invest Ophthalmol Vis Sci.,* vol. 59, no. 14, pp. 41-47, 2018.

[32] Q. Zhang, M. Chrenek, S. Bhatia, A. Rashid, S. Ferdous, K. Donaldson, H. Skelton, W. Wu, T. See, Y. Jiang, N. Dalal, J. Nickerson and H. Grossniklaus, "Comparison of histologic findings in age-related macular degeneration with RPE flatmount images.," *Mol. Vis.,* vol. 25, pp. 70-78, 2019.

[33] A. . London, I. . Benhar and M. . Schwartz, "The retina as a window to the brain—from eye research to CNS disorders," *Nature Reviews Neurology,* vol. 9, no. 1, pp. 44-53, 2013.

[34] C. L. Morgia, F. N. Ross-Cisneros, A. A. Sadun and V. . Carelli, "Retinal ganglion cells and circadian rhythms in Alzheimer's disease, Parkinson's disease, and beyond," *Frontiers in Neurology,* vol. 8, no. , pp. 1-8, 2017.

[35] W. Geeraets, R. Williams, G. Chan, W. Ham, D. Guerry and F. Schmidt, "The relative absorption of thermal energy in retina and choroid," *Investigative Ophthalmology & Visual Science,* vol. 3, no. 1, pp. 340-347, 1962.

[36] F. C. Delori and K. P. Pflibsen, "Spectral reflectance of the human ocular fundus," *Appl. Opt.,* vol. 28, pp. 1061-1077, 1989.

[37] J. D. Giese, T. N. Ford and J. Mertz, "Fast volumetric phase-gradient imaging in thick samples.," *Optics Express,* vol. 22, no. 1, pp. 1152-1162, 2014.

[38] D. . Carpentras, T. . Laforest, M. . Künzi and C. . Moser, "Effect of backscattering in phase contrast imaging of the retina," *Optics Express,* vol. 26, no. 6, pp. 6785-6795, 2018.

[39] D. T. Miller, O. P. Kocaoglu, Q. . Wang and S. . Lee, "Adaptive optics and the eye (super resolution OCT)," *Eye,* vol. 25, no. 3, pp. 321-330, 2011.

[40] International Organization for Standardization, *15004-2 Ophthalmic instruments — Fundamental requirements and test methods — Part 2: Light hazard protection,* Geneva, 2007.

[41] ICNIRP, "ICNIRP Guidelines on limits of exposure to incoherent visible and infrared radiation," *Health Physics,* vol. 105, no. 1, pp. 74-96, 2013.

[42] N. Geffen, S. Ofir, A. Belkin, F. Segev, Y. Barkana, A. Messas, E. Assia and M. Belkin, "Transscleral selective laser trabeculoplasty without a gonioscopy lens," *J Glaucoma,* vol. 26, no. 3, pp. 201-207, 2017.

[43] P. Thévenaz, U. E. Ruttimann and M. Unser, "A pyramid approach to subpixel registration based on intensity," *IEEE Transactions on Image Processing,* vol. 7, no. 1, pp. 27-41, 1998.

[44] N. Strouthidis, E. White, V. Owen, T. Ho, C. Hammond and D. Garway-Heath, "Factors affecting the test-retest variability of Heidelberg retina tomograph and Heidelberg retina

tomograph II measurements," *British Journal of Ophthalmology,* vol. 89, no. 11, pp. 1427-1432, 2005.

[45] Z. Monemhaghdoust, F. Montfort, Y. Emery, C. Depeursinge and C. Moser, "Off-axis digital holographic camera for quantitative phase microscopy," *Biomedical Optics Express,* vol. 5, no. 6, pp. 1721-1730, 2014.

[46] L. Tian and L. Waller, "Quantitative differential phase contrast imaging in an LED array microscope.," *Optics Express,* vol. 23, no. 9, pp. 11394-11403, 2015.

[47] S. B. Mehta and C. J. R. Sheppard, "Quantitative phase-gradient imaging at high resolution with asymmetric illumination-based differential phase contrast," *Optics Letters,* vol. 34, no. 13, pp. 1924-1926, 2009.

[48] D. Hillmann, H. Spahr, C. Hain, H. Sudkamp and G. Franke, "Aberration-free volumetric high-speed imaging of in vivo retina," *Scientific Reports,* vol. 6, no. 35209, 2016.

[49] N. D. Shemonski, F. A. South, Y. Z. Liu, S. G. Adie, P. S. Carney and S. A. Boppart, "Computational high-resolution optical imaging of the living human retina," *Nature Photonics,* vol. 9, no. 7, pp. 440-443, 2015.

[50] P. Xiao, V. Mazlin, K. Grieve, J.-A. Sahel, M. Fink and A.-C. Boccara, "In vivo high-resolution human retinal imaging with wavefront-correctionless full-field OCT," *Optica,* vol. 5, no. 4, pp. 408-412, 2018.

[51] J. Tam, L. Jianfei, A. Dubra and R. Fariss, "In Vivo Imaging of the Human Retinal Pigment Epithelial Mosaic Using Adaptive Optics Enhanced Indocyanine Green Ophthalmoscopy," *Investigative ophthalmology & visual science,* vol. 57, no. 10, pp. 4376-84, 2016.

[52] R. H. Masland, "The Neuronal Organization of the Retina," *Neuron,* vol. 76, no. 2, pp. 266-280, 17 October 2012.

[53] F. Zernike, "Phase contrast, a new method for the microscopic observation of transparent objects part II.," *Physica,* vol. 9, no. 19, pp. 974-986, 1942.

[54] L. Tian, J. Wang and L. Waller, "3D differential phase-contrast microscopy with computational illumination using an LED array," *Optics Letters,* vol. 39, no. 5, pp. 1326-1329, 2014.

[55] B. . Rappaz, P. . Marquet, E. . Cuche, Y. . Emery, C. . Depeursinge and P. J. Magistretti, "Measurement of the integral refractive index and dynamic cell morphometry of living cells with digital holographic microscopy," *Optics Express,* vol. 13, no. 23, pp. 9361-9373, 2005.

[56] G. Nomarski, "Microinterféromètre différentiel à ondes polarisées," *Journal de Physique et le Radium,* vol. 16, pp. 9S-11S, 1955.

[57] P. Marquet, B. Rappaz, P. J. Magistretti, E. Cuche, Y. Emery, T. Colomb and C. Depeursinge, "Digital holographic microscopy: a noninvasive contrast imaging technique allowing quantitative visualization of living cells with subwavelength axial accuracy," *Optics Letters,* vol. 30, no. 5, p. 468–470, 2005.

[58] Z. Liu, L. Tian, S. Liu and L. Waller, "Real-time brightfield, darkfield, and phase contrast imaging in a light-emitting diode array microscope.," *Journal of Biomedical Optics,* vol. 19, no. 10, pp. 106002-106002, 2014.

[59] J. Kühn, E. Shaffer, J. Mena, B. Breton, J. Parent, B. Rappaz, M. Chambon, Y. Emery, P. Magistretti, C. Depeursinge, P. Marquet and G. Turcatti, "Label-free cytotoxicity

screening assay by digital holographic microscopy," *Assay Drug Development Technologies,* vol. 11, no. 2, pp. 101-107, 2013.

[60] R. S. Jonnal, O. P. Kocaoglu, R. J. Zawadzki, Z. Liu, D. T. Miller and J. S. Werner, "A Review of Adaptive Optics Optical Coherence Tomography: Technical Advances, Scientific Applications, and the Future," *Investigative Ophthalmology & Visual Science,* vol. 57, no. 9, pp. 51-68, 2016.

[61] J. G. Fujimoto, B. Bouma, G. J. Tearney, S. A. Boppart, C. Pitris, J. F. Southern and M. E. Brezinski, "New Technology for High-Speed and High-Resolution Optical Coherence Tomography," *Annals of the New York Academy of Sciences,* vol. 838, no. 1, pp. 97-107, 1998.

[62] T. N. Ford, K. K. Chu and J. Mertz, "Phase-gradient microscopy in thick tissue with oblique back-illumination," *Nature Methods,* vol. 9, no. 12, pp. 1195-1197, 2012.

[63] W. Drexler and J. G. Fujimoto, "State-of-the-art retinal optical coherence tomography," *Progress in Retinal and Eye Research,* vol. 27, no. 1, pp. 45-88, 2008.

[64] B. Lumbroso, D. Huang and M. Rispoli, Angio OCT in Everyday Ophthalmic Practice, Jaypee Brothers Medical Publishers, 2017.

## Acknowledgments

The authors thank Dr Michaël Nicolas from the eye bank of the Jules-Gonin Eye Hospital for providing a *post-mortem* human eye, Dr. Irmela Mantel together with the team of the center for clinical investigation for their time spent in performing the ophthalmologic checks for our study on healthy participants, and Dr. Sylvain Roy, Dr. Alexandre Matet and Dr. Daniel Sage for the fruitful discussions.

## Funding

In addition to the research partners, this study was supported by the following programs: Enable program of the Technology Transfer Office at EPFL, EPFL Innogant, Bridge proof of concept (InnoSuisse and SNSF), Gebert Rüf Stiftung foundation (GRS-052/17) and EIT Health Innovation by idea (19323-ASSESS).

## Author contributions

TL designed and built the ex vivo microscope, obtained the ex vivo results, built the in vivo device, wrote the code for ex-vivo and in vivo processing, and wrote the paper. DC developed the theoretical model and wrote the paper. MK designed and built the in vivo device, wrote the code for in vivo processing, and wrote the paper. LK provided and prepared the ex vivo samples, participated in the interpretation to the phase images, and wrote the paper. FBC supervised the project and wrote the paper. CM designed the experiment, supervised the project, and wrote the paper.

## Competing interests

## Data and materials availability

All data needed to evaluate the conclusions in this paper are present in the paper and/or the Supplementary Materials. Additional data related to this paper may be requested from the authors.

## Figures and Tables

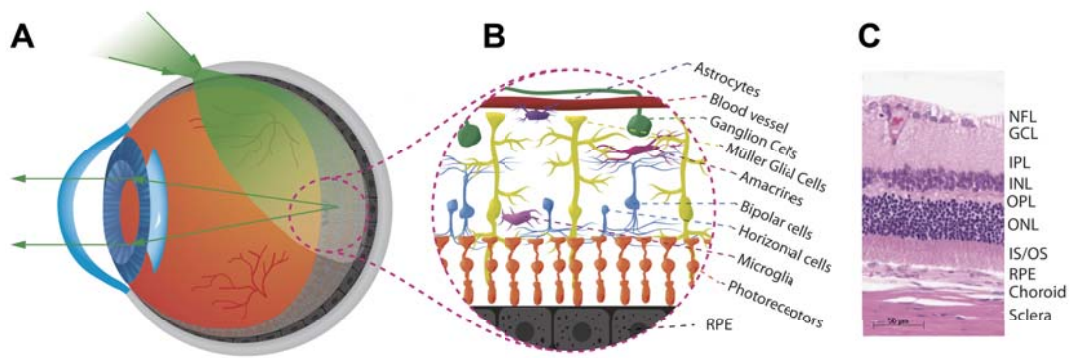

**Fig. 1: Illumination of the retinal layers provided via transscleral illumination.** (A) Light transmission into the eye. The light is first transmitted through the sclera, the retina pigment epithelium (RPE), and the neuroretina. After travelling through the vitreous humor and the retina, it impinges on the RPE layer. Here, by backscattering off, the RPE generates a new illumination beam. This secondary illumination provides a transmission light that propagates through the transparent media, i.e., the neuroretina and its translucent cells, the vitreous, the eye lens, and, then, the anterior segment of the eye. (B) Illustration of the translucent vascular, neuronal and glial cells of the retina. (C) Histological cut of the retinal layers, the choroid, and the sclera. GCL: ganglion cells layer; INL: inner nuclear layer; IPL: inner plexiform layer; IS/OS: inner and outer segments of the photoreceptors; NFL: nerve fiber layer; ONL: outer nuclear layer; OPL: outer plexiform layer.

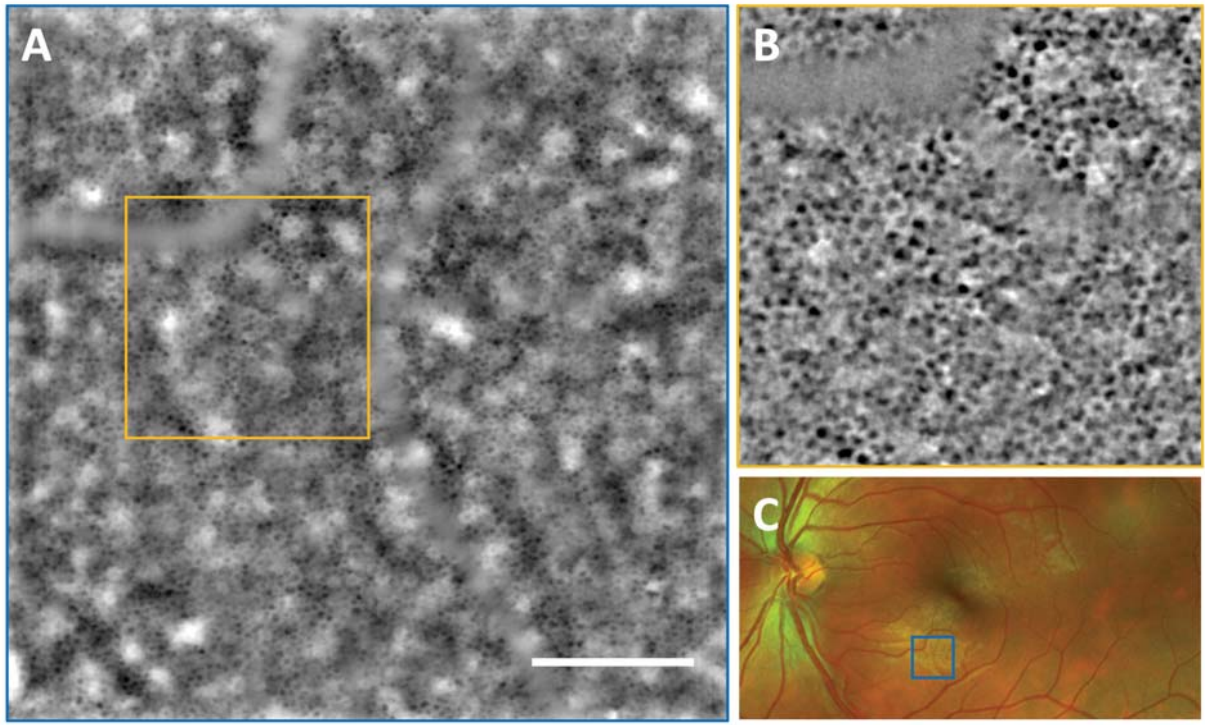

**Fig. 2.** *In vivo* **full-field (4.4°x4.4°) image of the retinal pigment epithelium (RPE) of a healthy volunteer taken with TOPI. The image of the left eye of the subject S9 was taken at an eccentricity of 6.7° from the fovea.** **(A)** The RPE image is the average of 100 raw retinal images taken in 9 seconds to increase the Signal over Noise Ratio. A nice mosaic of RPE cells is visible where the individual cell somas appear dark due to the absorbing pigment within the cell body. The cells are partially masked under a lower spatial frequency background modulation with varying intensity. This modulation is coming from the non-uniform pigmentation of the epithelium and the out-of-focus signal from the choroid. Scale bar is 1° or 300 μm. **(B)** The original RPE image is cropped and high-pass filtered to remove the out of focus light and have a more uniform RPE cell mosaic **(C)** SLO fundus image showing the location of the TOPI image.

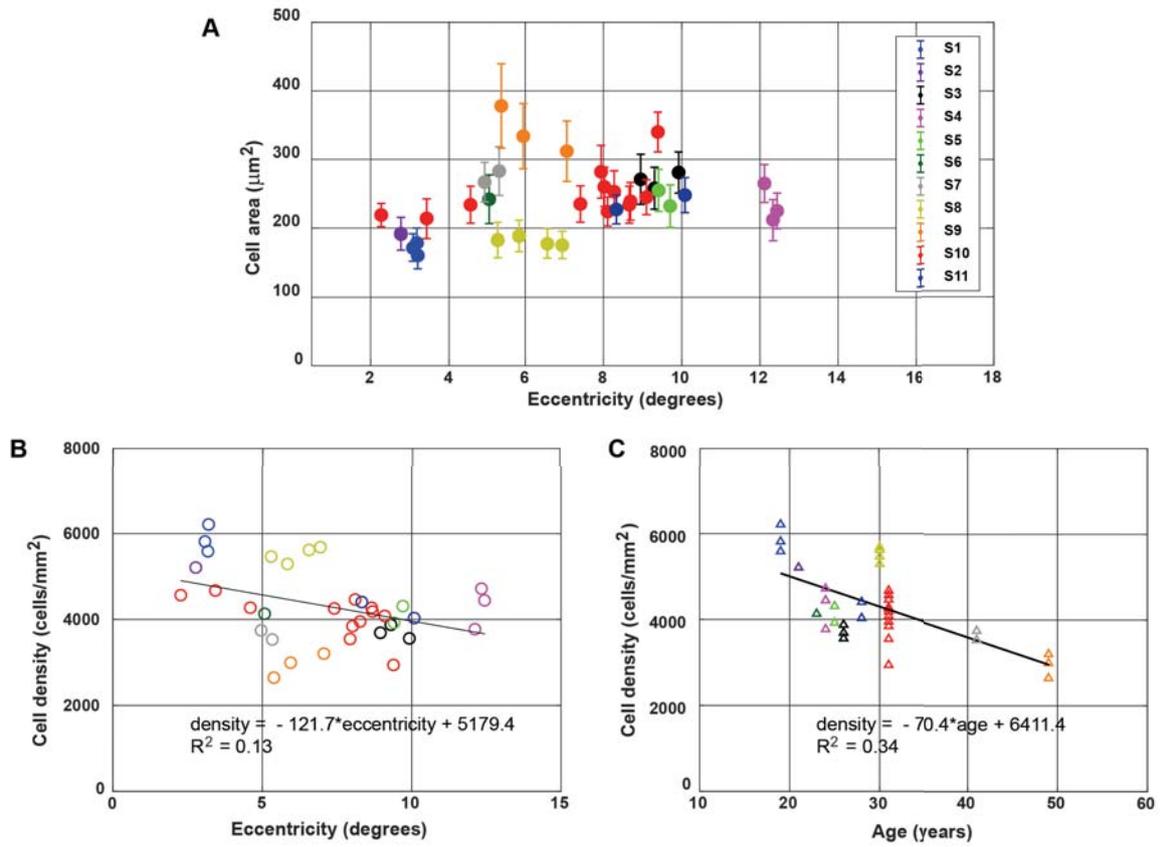

**Fig. 3.** *In vivo* **retinal pigment epithelium TOPI quantification of eleven healthy volunteers (S1 to S11). (A)** Quantification of RPE cell area versus the eccentricity with respect to the fovea. Quantification of RPE cell density versus the eccentricity with respect to the fovea **(B)** and age of the participants **(C).** The phototype of the subjects ranged from 1 to 6 on the Fitzpatrick scale **[19]**.

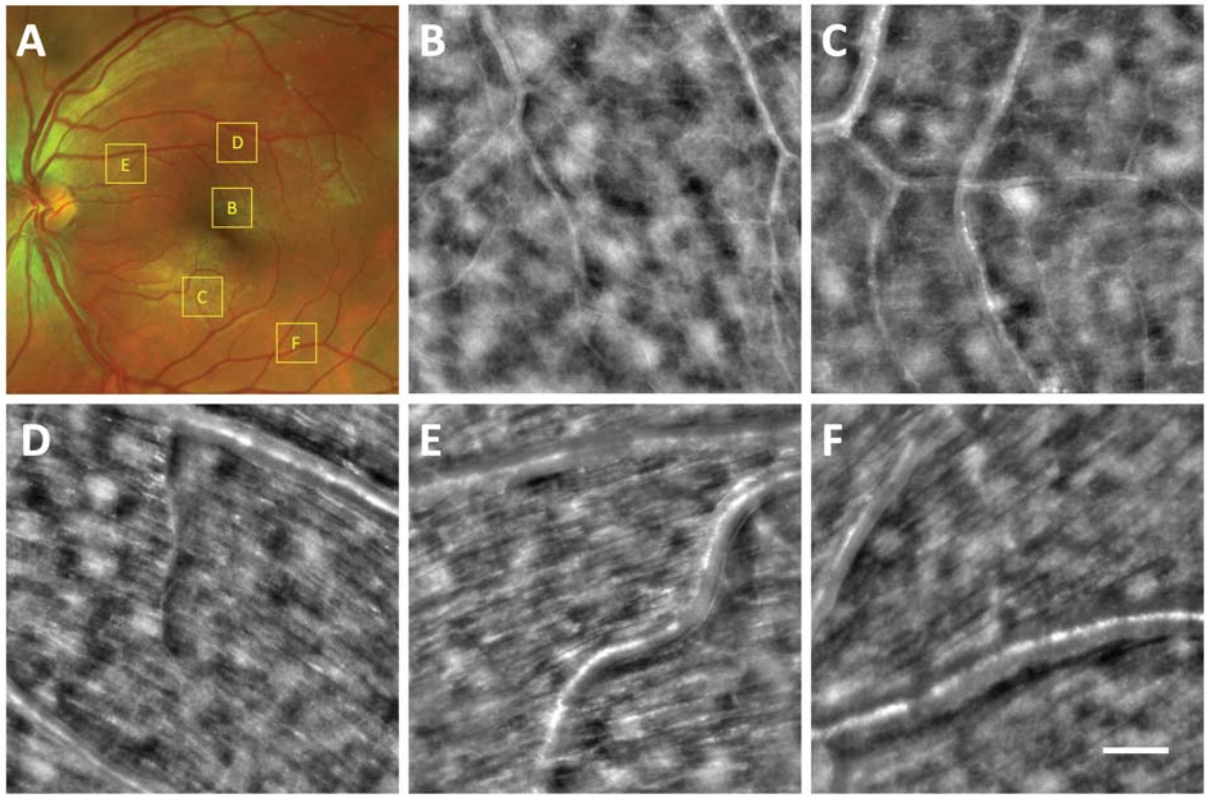

**Fig. 4: Microcapillaries and nerve fiber layer TOPI images of subject S10 taken at different eccentricities.** Each cropped image has a 3°x3° Field Of View, scale bar is 0.5° or 150 μm. **(A)** SLO fundus image showing the location of each TOPI image crop. **(B-C)** Microcapillaries being few microns thick are visible even if no blood flow is present thanks to the dark field effect of the oblique illumination. **(D-E-F)** Large axons bundles of 30-50 μm are clearly identifiable when looking at the nerve fiber layer. The bundles are well aligned and directed toward the optical nerve, as expected.

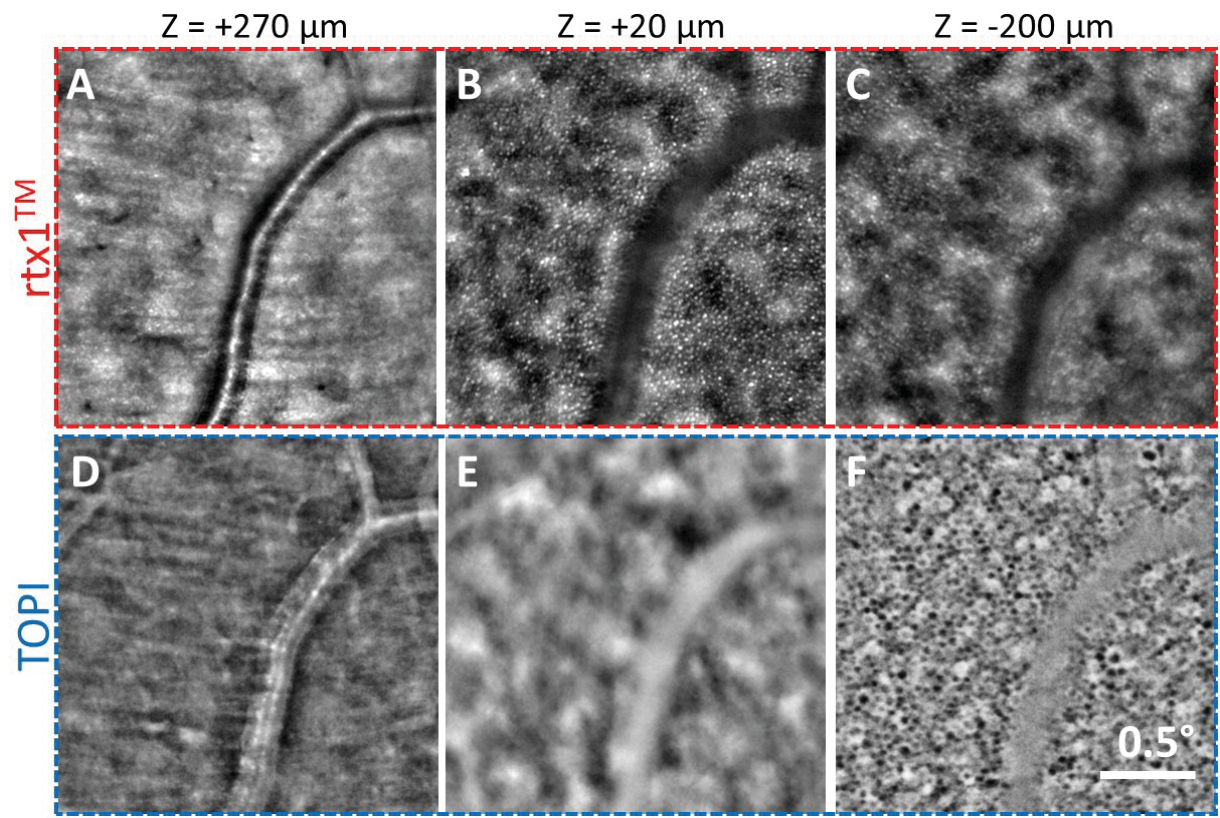

**Fig. 5: Comparison between** *in vivo* **images from a trans-pupil flood illumination commercial system (rtx1[TM], Imagine Eyes; Top) and a temporal trans-scleral illumination TOPI image (Bottom). (A, D)** At the nerve fiber layer level (z=270 µm), the shape of nerve is detected with both modalities. **(B, E)** At the photoreceptor level (z=0 µm), the cones are well visible with high contrast and appear hyper-reflective in the rtx1 image due to light coupling inside the cells (Stiles-Crawford effect). They are barely visible in the TOPI image because no reflection is produced by the oblique illumination. **(C, F)** At the RPE level (z=200 µm), on the rtx1 trans-pupil image, no RPE cells are visible because their signal is buried below the high reflection coming from the out-of-focus photoreceptors, whereas, on the TOPI image, a nice RPE cell mosaic is visible. The somas appear dark because of the pigments composing the cells.

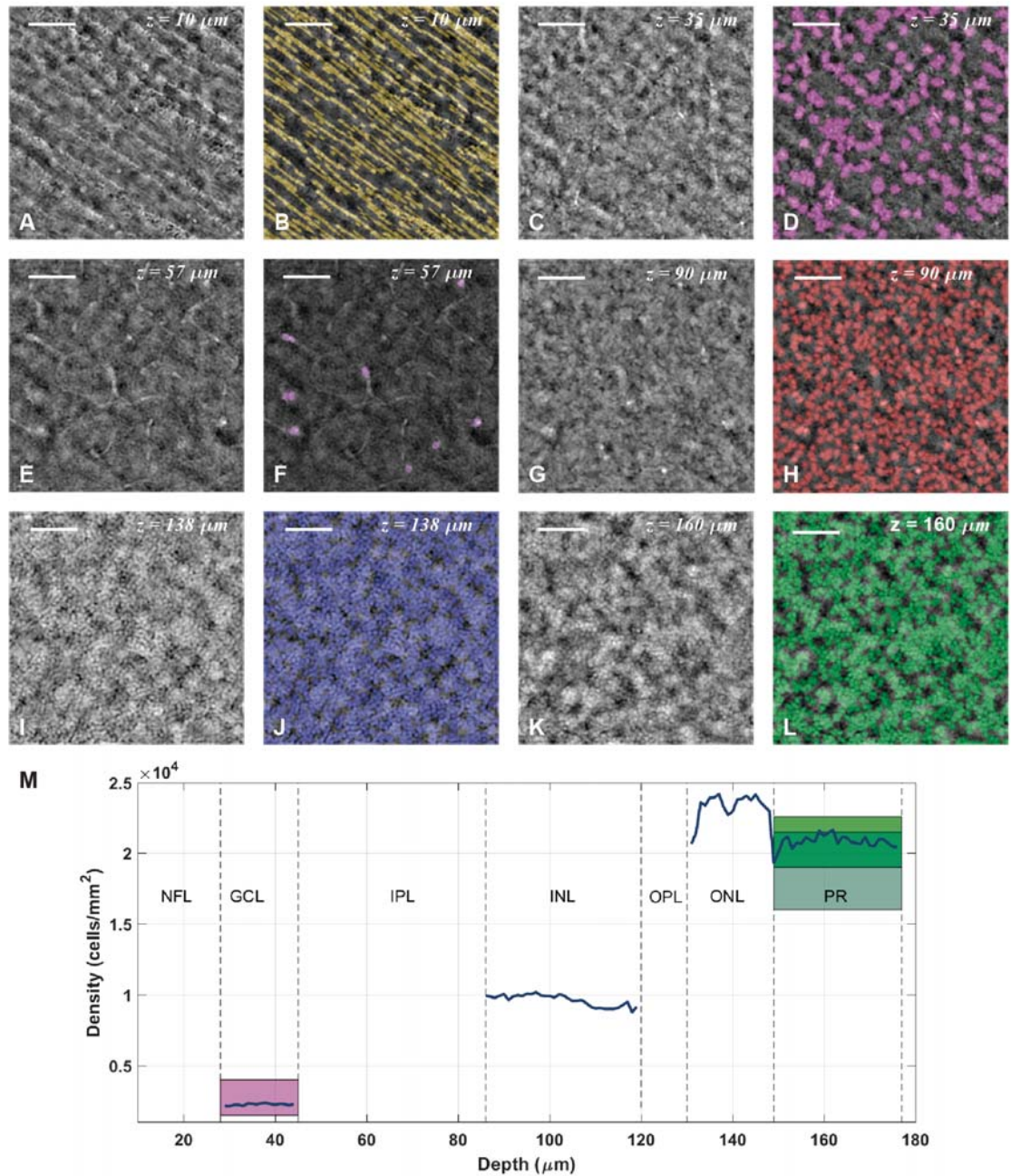

**Fig. 6: (A – L) In-depth scan of a flat-mounted fixed pig retina-choroid complex obtained via phase imaging. (A, C, E, G, I, K) Phase images**. (B, D,F,H, J, L) Phase images with digitally tagged cells and structures. Nerve fiber layer, z = 10 μm. Ganglion cell layer, z = 35 μm. Inner plexiform layer, z = 57 μm. Inner nuclear layer, z = 90 um. Outer nuclear layer, z = 138 μm. Inner / outer photoreceptors segments interface, z = 160 μm. Scale bars = 50 μm. **(M) Cell densities versus depth corresponding to A – L. Blue curves**: measured data in this study. **Magenta**: density range of ganglion cells obtained from **[25]**. **Green**: density range of cones obtained from **[26]**. **Dark green**: density range of cones obtained from **[24]**. GCL: ganglion cell layer; INL: inner nuclear layer; IPL: inner plexiform layer; NFL: nerve fiber layer; ONL: outer nuclear layer; OPL: outer plexiform layer; PR: interface between the inner and outer segments of the photoreceptors.

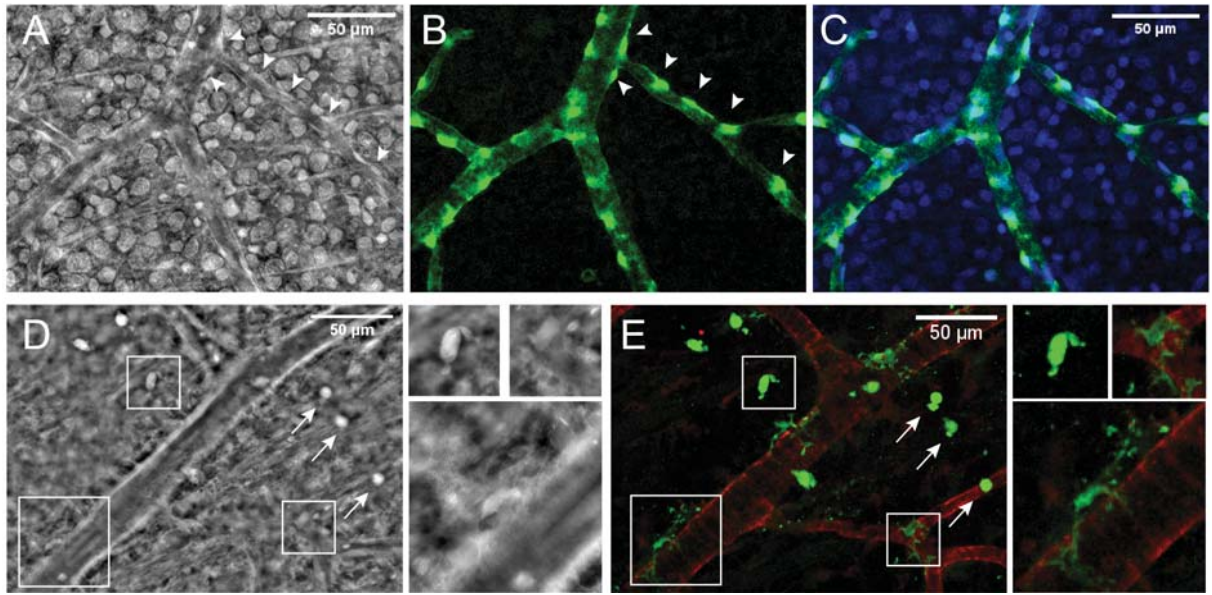

**Fig. 7: Image correlation analysis between high resolution phase microscopy and fluorescence confocal microscopy of retinal vessels and their surrounding cells. (A-C) Pericytes visualization in retina from a healthy rat aged 3 months.** Retinal phase images (A) were compared with fluorescence microscopy images of the retina immunostained for NG2 (green in B and C) and counter-stained with DAPI (blue in C). Cells highlighted with white arrowheads in the phase image were localized exactly where the pericytes were detected using fluorescence confocal microscopy. **(D-E) Microglial cells visualization in retina from an 11-month-old rat presenting retinal dystrophy associated with telangiectasias.** Retinal phase images (D) were compared with fluorescence microscopy images of the retina immunostained with antibodies against Iba1 (green in E) and collagen IV (red in E). Magnifications of three cells in both imaging modalities exemplified the detection of microglial cells using high resolution phase imaging. Ramified q**uiescent microglia as well as** reactive **microglia**, with a round cell body **(ar**r**ows)**, **were observed**.

# Supplementary Materials

|  | OCT with CAO[1] | Full field OCT | AO-OCT | AO-SLO[2] | AO flood illumination | AO TOPI |
|---|---|---|---|---|---|---|
| **Axial resolution** | 4.5 µm | 8 µm | 2 µm | 50 µm | 50 – 100 µm | 50 – 100 µm |
| **Lateral resolution** | 2 – 3 µm | 4 µm | 3 µm | 2 µm | 3 µm | 3 µm |
| **Field Of View (FOV) (°)** | 1.5 × 1.5 | 2.4 × 2.4 | 1.5 × 1.5 | 2 × 2 | 4 × 4 | 4.4 × 4.4 |
| **System FOV (°)** | NA | NA | NA | 30° | 30° | 30° |
| **Digital sampling (µm/pixel)** | 1 – 2 | 0.5 | 1 | 0.67 – 2 | 0.8 | 0.7 |
| **Time for full retina scan** | 10 – 20 sec | N/A | 15 min | N/A | N/A | 1 min 20 s |
| **Time for NFL/GCL imaging** | 10 – 20 sec[3] | 0.2 sec | 15 min[3] | ~10 sec | 2 sec | 9 sec |
| **Time for RPE imaging** | N/A | N/A | 15 min[3] | 12 sec | N/A | 2 sec |
| **Time for cone imaging** | 10 – 20 sec[3] | 0.2 sec | 15 min[3] | ~10 sec | 2 sec | 9 sec |
| **Nerve Fiber Layer (NFL)** | yes | yes | yes | yes | yes | yes |
| **Ganglion Cell Layer (GCL)** | no | no | yes | yes | no | yes |
| **Cones** | yes | yes | yes | yes | yes | yes |
| **Foveal cones** | no | no |  | yes |  | no |
| **Rods** | no | no |  | yes | no | no |

| Retinal Pigment Epithelium (RPE) | no | no | yes | yes | no | yes |
|---|---|---|---|---|---|---|
| Ref/Company | [48], [49] | [50] | [15], [18] | Boston Micromachines Corp., PSI Inc, [14], [10], [8], [51] | Imagine Eyes, [22] | Present study |

Table 1: **Comparison of the proposed method with state of the art commercial devices or laboratory prototypes**. [1] Computational adaptive optics. [2] Includes autofluorescence, confocal, split detector and offset aperture SLO systems. [3] The limiting speed is the x,y scanning speed.

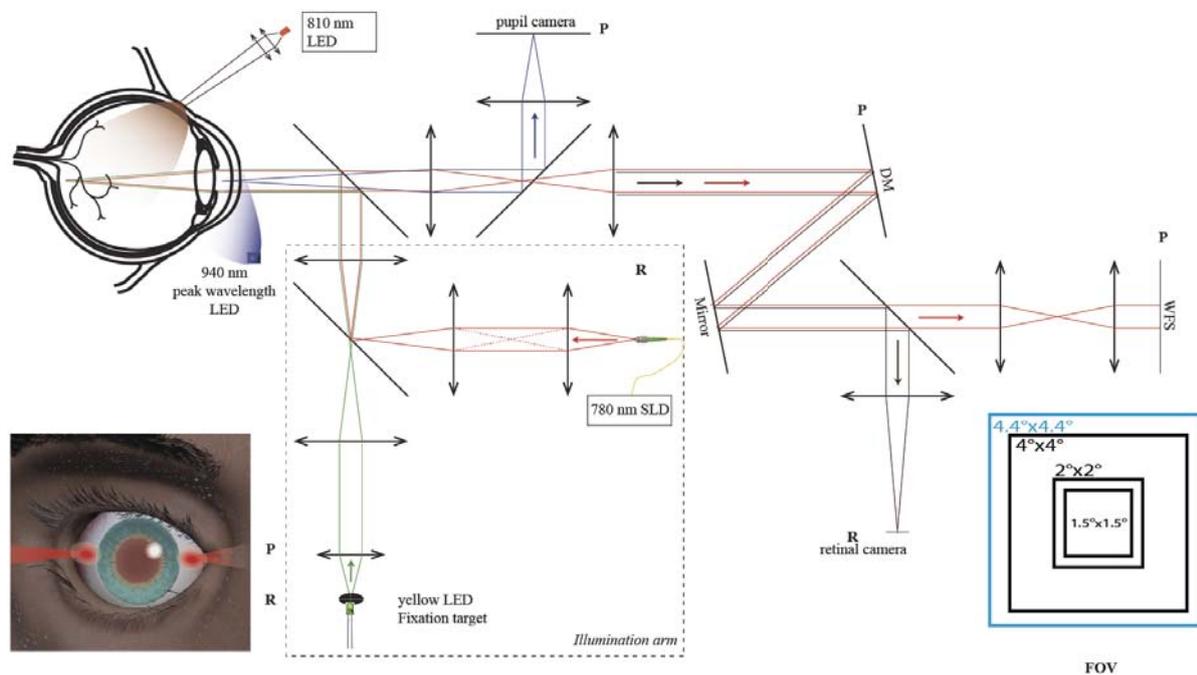

**Fig. S 1. Schematic of our *in vivo* phase imaging system.** The digital photography represents transscleral illuminations by means of focused beams on the sclera. Light is then transmitted inside the eyeball. After scattering off the eye fundus, the light passing through the retinal cell layers is collected through the pupil of the eye. The optical setup includes an adaptive optics loop to correct the aberration of the eye. DM: Deformable mirror; LED: light-emitting diode; SLD: superluminescent diode; WFS: wavefront sensor. At the bottom right is a representation of the fields of view obtained via state-of-the-art methods (black, see Table 1 for correspondence) and TOPI (blue).

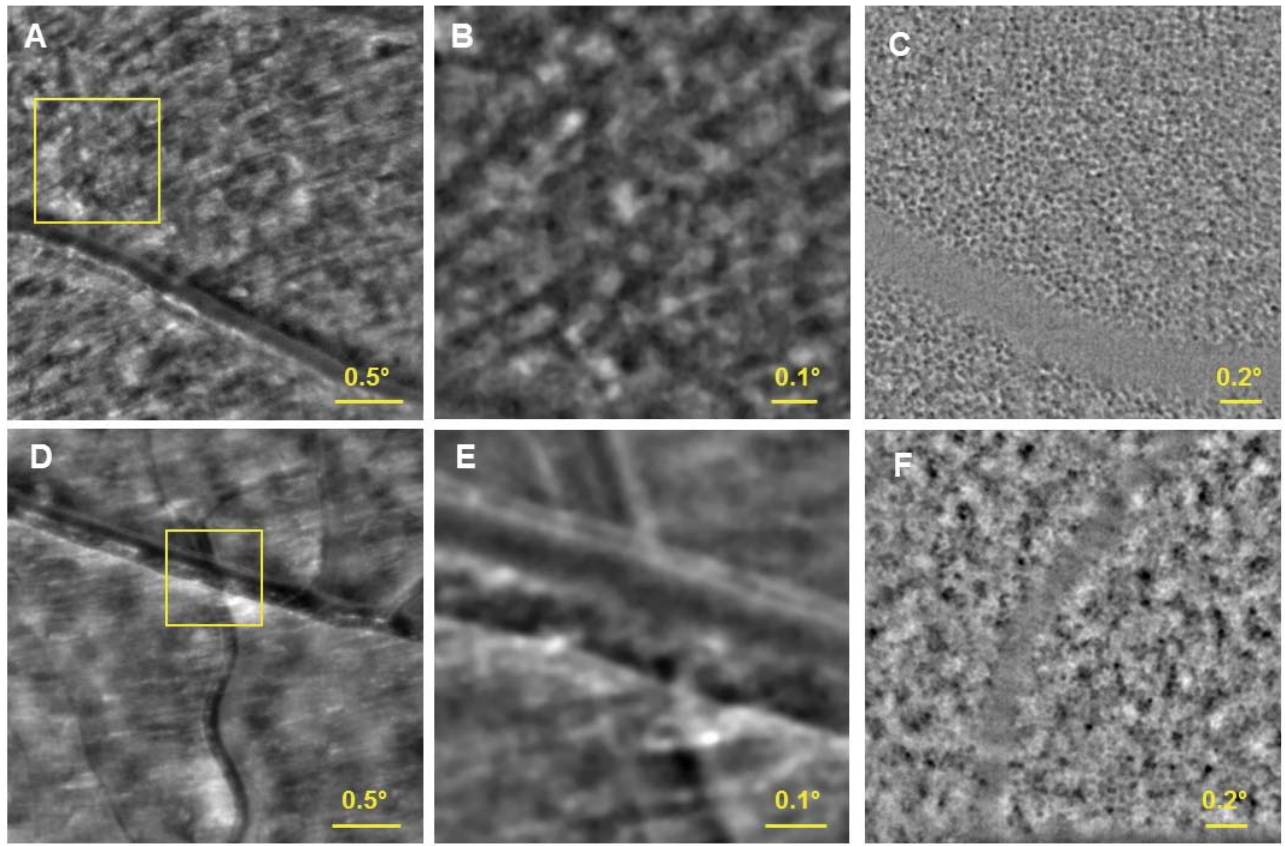

**Fig. S 2.** *In vivo* **TOPI images of two distinct retinal layers of two healthy volunteers S4 (12° eccentricity from the fovea) and S11 (8° and 5° eccentricity from the fovea). (A, D)** ganglion cells layers (z=200 μm), **(B, E)** small crops showing the yellow region of interest on **A** and **D**, **(C, F)** retinal pigment epithelium (z=0 μm),.

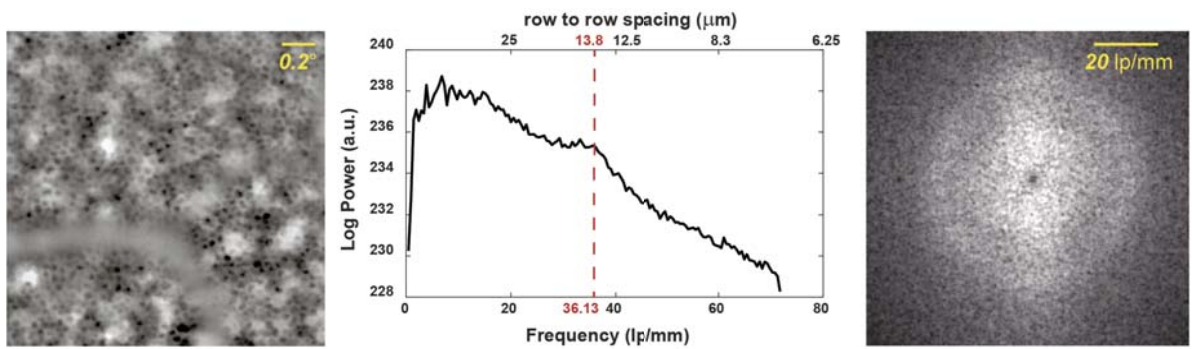

**Fig. S 3. Fourier analysis of an RPE layer image centered at 7° from the fovea. (Left)** Dark-field image that was high-pass filtered to enhance the visualization. **(Center)** Axial profile of the spectrum. **(Right)** Spectrum of the left image cropped at a row to row spacing of 6.8 μm. The peak signature of the RPE signal was located at 13.8 μm, which is consistent with the values found in the literature.

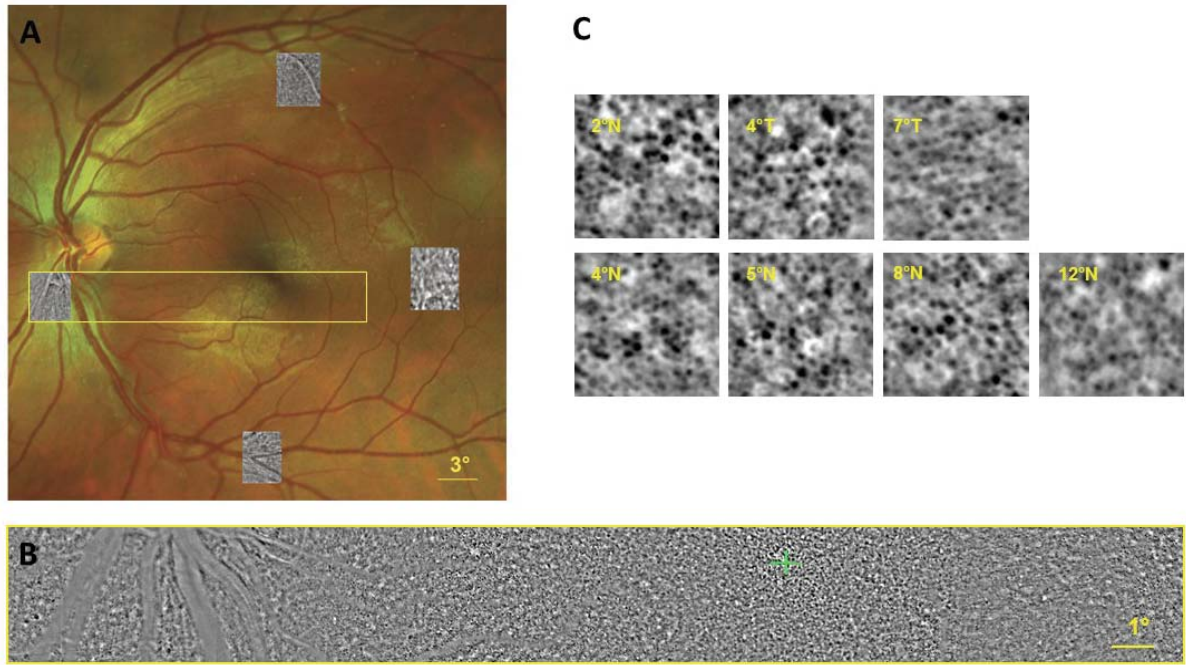

**Fig. S 4: Representation of the maximum eccentricities of our *in vivo* TOPI device. (A)** Subject S9, left eye fundus as seen using a scanning laser ophthalmoscope (Optos). *In-vivo* TOPI at NFL level placed on top of the large FOV SLO image. Subject S10 left eye. On the vertical axis, TOPI images are located at ±15°. On the horizontal axis, images are located at ±15°. The yellow rectangle illustrates the wide field of view stitched to obtain a 27.4° × 3.5° image shown in (**B**, *full resolution image available*). The green cross indicates the fovea. **(C)** TOPI crop images of 0.5° × 0.5° taken at the RPE level at different eccentricities. The individual RPE cells appear in dark with bright edges and are clearly identifiable on most part of the image. Note that the RPE cells below the retinal vessels are hidden.

Page **40** of **45**

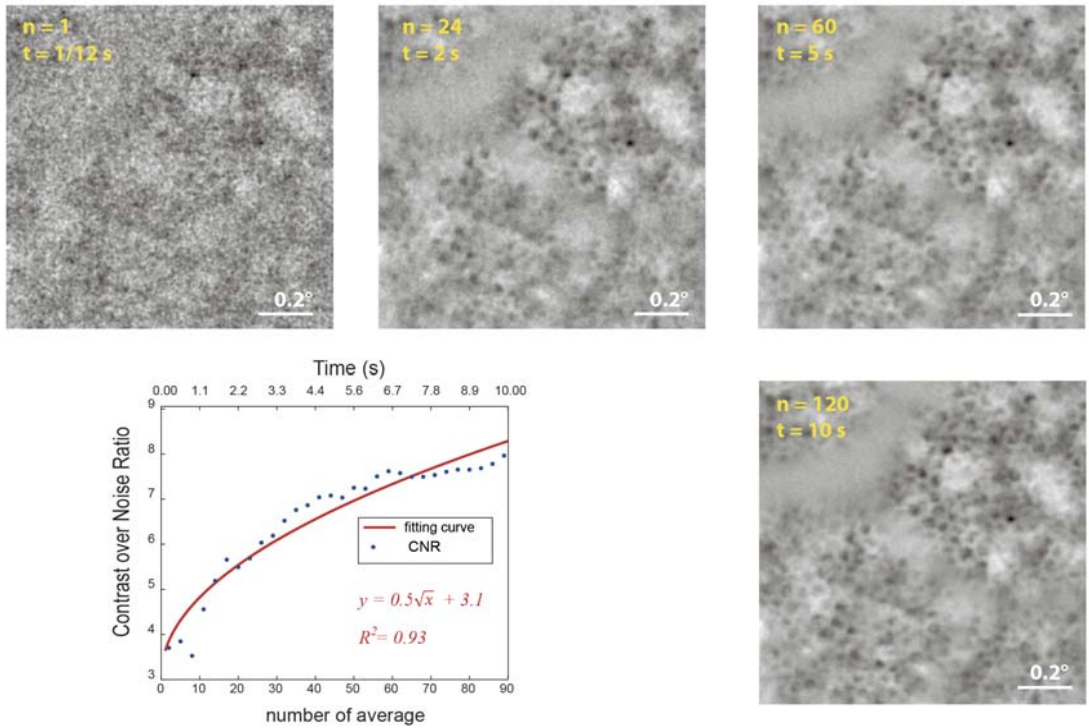

**Fig. S 5. Averaging of the retinal pigment epithelium layer image centered at 7° from the fovea. (Top, Bottom right)** Phase images were acquired from two illumination points at 12 Hz. The dark-field images were aligned before averaging. **(Bottom Left)** Impact on the SNR based on the power spectrum analysis.

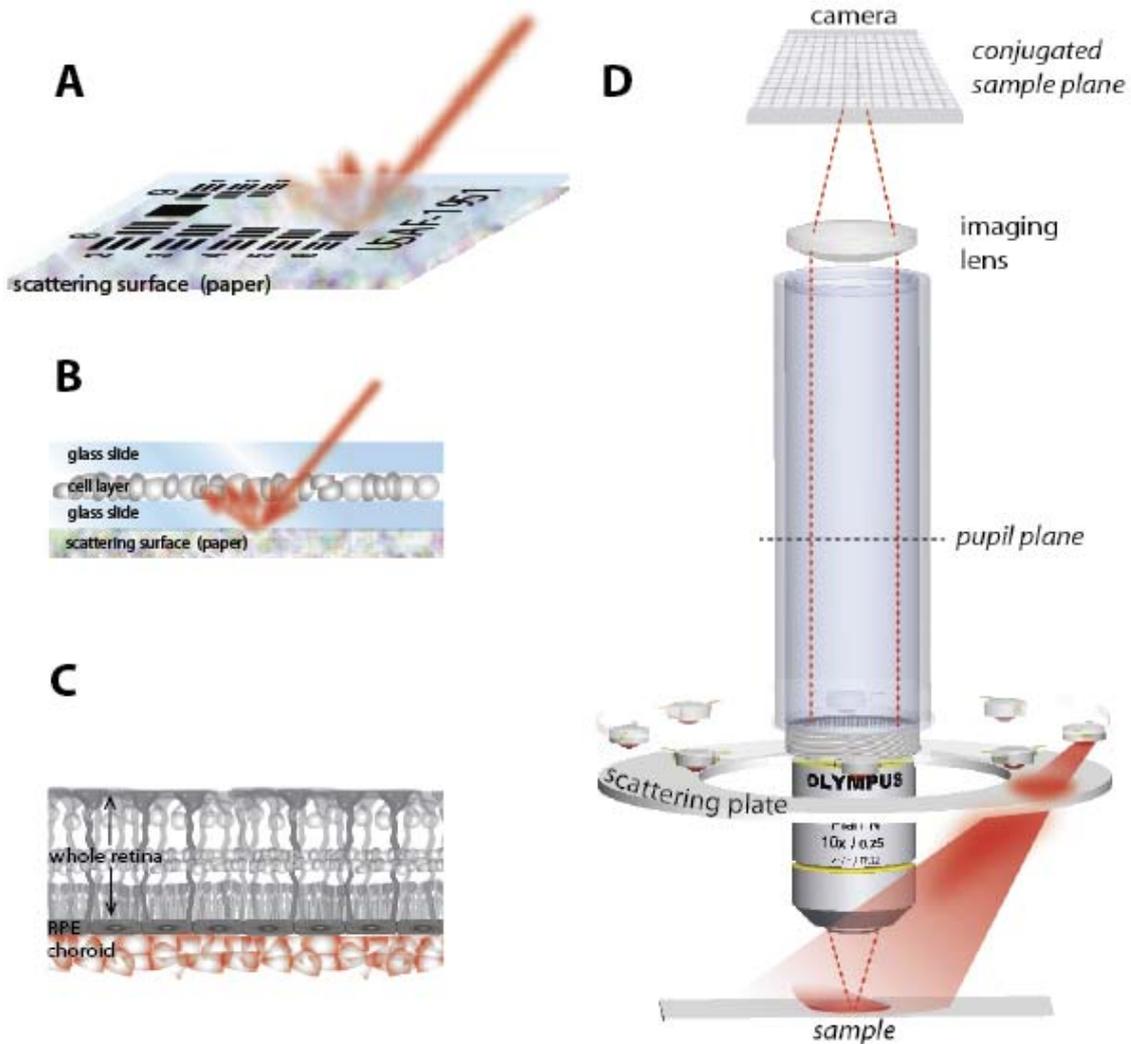

**Fig. S 6. Experimental set up for *ex vivo* observations. (A–C) Schematics of the imaged samples**. **(A)** Intensity USAF target sample mounted on top of a scattering surface (paper) for assessing the modulation transfer function of our optical system. Incoming light back-scatters off the paper. **(B)** Frontal frozen sections of a pig neuroretina, mounted on a glass slide. A scattering surface (paper) is attached to the bottom side of the glass slide in order to provide back-scattering light. **(C)** Flat-mounted "retina - choroid" complex placed on a glass slide. The light back-scatters from the RPE-choroid layer **(D) Experimental microscope for phase imaging.** LED light illuminates the sample from multiple directions. The backscattered light effectively becomes a forward beam that illuminates the transparent phase layers in the transmission geometry before reaching the microscope objective (10×, 0.25NA). An imaging lens then makes an image of the sample on a camera.

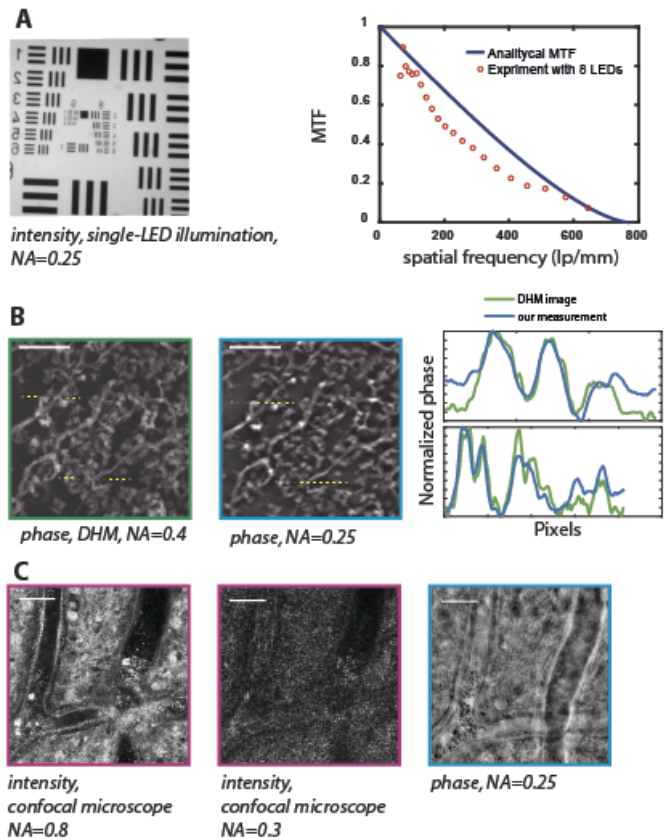

**Fig. S 7:** *Ex vivo* **imaging of (A) the USAF intensity target, (B) a pig neuroretina section, and (C) a human flat-mounted retina-choroid complex.** (A) Intensity image of the USAF intensity target. (Left) Dark-field image of the USAF intensity target obtained via single-LED illumination. (Right) Plot of the theoretical and experimental modulation transfer functions obtained from the image. Red dots correspond to the contrasts of the elements of groups 8 and 9 of the USAF target. (B) Images of a 10-μm-thick frontal frozen section of pig neuroretina at the inner nuclear layer level. (Left) Phase image obtained using a digital holographic microscope. (Center) Phase image obtained using the proposed method. (Right) Horizontal cross section plot of quantitative phase comparison. Scale bars = 30 μm. (C) Images of a human flat-mounted fixed retina-choroid complex at the nerve fiber layer level. (Left-middle) Reflectance confocal images taken with 0.8-NA and 0.3-NA objectives, respectively. (Right) Phase image taken with a 0.25-NA objective. Scale bars = 50 μm.

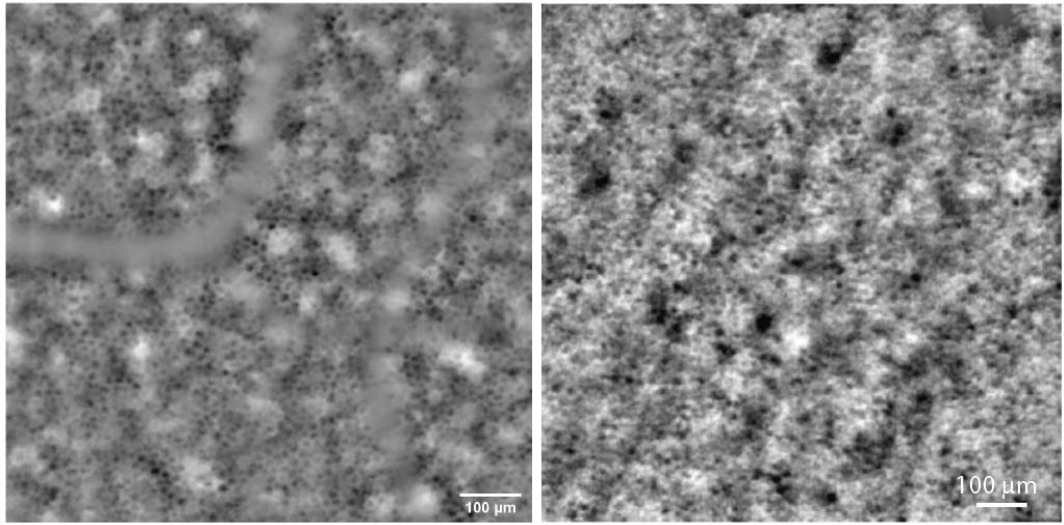

**Fig. S 8. Comparison of in *vivo* and *ex vivo* dark-field images of the RPE layer. (Left)** *In vivo* human RPE dark-field image acquired with a ~5 mm diameter pupil and illuminated using a LED with an 810-nm peak wavelength. **(Right)** *Ex vivo* human RPE registered with a NA of 0.15 and illuminated using an LED with an 830-nm peak wavelength.

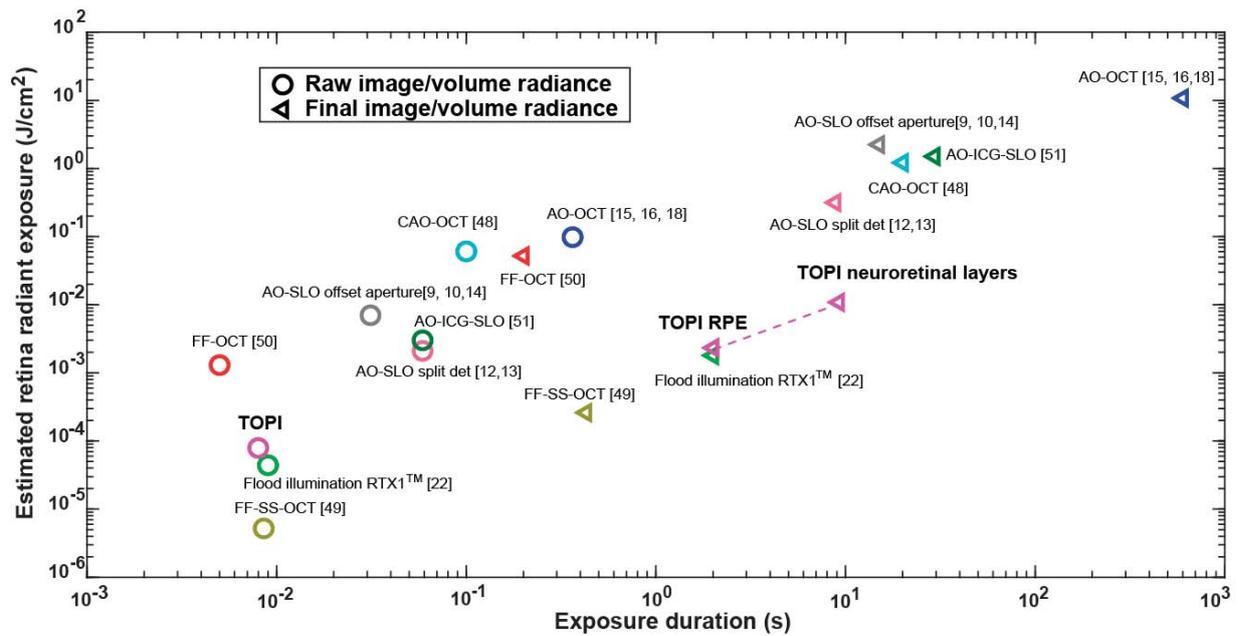

**Fig. S 9. Estimated radiant exposure at the imaged retinal area. Circles** show the energy for raw image or volume acquisition, while triangles show the energy needed to obtain a final image resulting from the average of several raw images. AO: Adaptive optics; CAO: Computational AO; ICG: Indocyanine Green; FF: Full Field; OCT: Optical Coherence Tomography; SLO: Scanning Laser Ophthalmoscopy; SS: Swept Source; TOPI: Transcleral Oblique Phase Imaging.

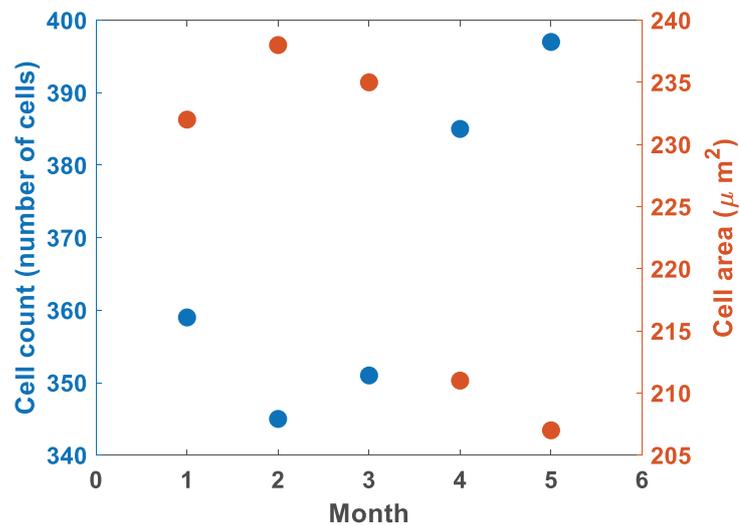

**Fig. S 10.** *In-vivo* RPE cells count and area over 5 month at 4° eccentricity for a 0.96°×0.96° region of interest, on subject S10.